\documentclass[twocolumn,american,floatfix, table, dvipsnames,superscriptaddress]{revtex4-1}
\usepackage[T1]{fontenc}
\usepackage[latin9]{inputenc}
\usepackage{color}
\usepackage{babel}
\usepackage{amsmath}
\usepackage{amssymb}
\usepackage{graphicx}
\usepackage{esint}
\usepackage[unicode=trueam,
 bookmarks=false,
 breaklinks=true,pdfborder={0 0 1},backref=false,colorlinks=true]
 {hyperref}
\hypersetup{
 pdfcreator={},pdfproducer={LaTeX with hyperref},linkcolor=blue,anchorcolor=blue,citecolor=blue,filecolor=red,menucolor=red,pagecolor=red,urlcolor=blue,pdfstartview=FitV,pdfhighlight=/I,pdfpagelayout=TwoColumns,hypertexnames=true}

\makeatletter
\usepackage{lmodern}
\usepackage{mathptmx, newtxtext, newtxmath, xspace}
\usepackage{bbm}

\usepackage[table, dvipsnames]{xcolor}
\definecolor{colorA}{cmyk}{0,0,0,0.05}
\definecolor{colorB}{cmyk}{0.14,0.04,0,0}
\definecolor{colorC}{cmyk}{0.02,0.0799,0,0}
\definecolor{colorD}{cmyk}{0.099,0.14,0,0}
\newcommand{\hhbar}{\mathchar'26\mkern-9mu h}
\usepackage{natbib}
\setcitestyle{numbers}

\makeatother

\begin{document}
\title{Strong-coupling superconductivity near Gross-Neveu quantum criticality
in Dirac systems }
\author{Veronika C. Stangier }
\affiliation{Institute for Theory of Condensed Matter, Karlsruhe Institute of Technology,
Karlsruhe 76131, Germany}
\author{Daniel E. Sheehy }
\affiliation{Department of Physics and Astronomy, Louisiana State University, the
Baton Rouge, LA 70803 USA}
\author{J{\"o}rg Schmalian }
\affiliation{Institute for Theory of Condensed Matter, Karlsruhe Institute of Technology,
Karlsruhe 76131, Germany}
\affiliation{Institute for Quantum Materials and Technologies, Karlsruhe Institute
of Technology, Karlsruhe 76131, Germany}
\date{\today }
\begin{abstract}
 We study two-dimensional massless Dirac fermions at neutrality, coupled to bosonic modes through a Yukawa interaction. We then examine the intriguing possibility that such a system, devoid of carriers
at zero temperature, might nevertheless exhibit superconductivity. Remarkably, we find that superconductivity emerges in the vicinity of Gross-Neveu quantum criticality, provided the fermions cease to behave as well-defined quasiparticles, that is, once their anomalous dimension in the normal state becomes sufficiently large. In other words, well-defined fermions do not superconduct, whereas ill-defined ones do.
We analyze four symmetry-distinct bosonic modes, each capable of driving normal-state criticality and, in three of the four cases, giving rise to a distinct superconducting phase. While phase fluctuations are strong in this regime, we argue that they do not destroy the superconducting state.  We further characterize the resulting pairing states for a concrete Dirac model of spin-orbit coupled systems with orbitals of different parity. Our results are obtained using the SYK-inspired framework for Dirac systems introduced by Kim et al.~\cite{Kim2021}, which provides a controlled approach to the strongly coupled regime of Dirac fluids near Gross-Neveu criticality.
\end{abstract}
\maketitle

\section{Introduction}

The exploration of two-dimensional gapless Dirac materials~\cite{Wehling2014,Vafek2014,Boyack2021}, i.e. systems
in which low-energy excitations mimic relativistic,  Dirac
fermions, is of importance in several prominent materials such as
single layer graphene~\citep{novoselov2004, castro2009}, twisted bilayer graphene~\citep{li2010,cao2018,cao2018b,andrei2020}, twisted double-layer WSe$_{2}$~\cite{Angeli2021,Pan2023,Foutty2023,Ma2024,Tolosa2025},
surface states of three-dimensional topological insulators~\cite{Hasan2010,Qi2011}, near-ferroelectric semimetals~\cite{Kozii2019,Kozii2022}, or in the case of merging Dirac points~\cite{Montambaux2009,Isobe2015,Cho2016novel,Link2018}.
Near the neutrality point of a massless Dirac fermion, the density
of states at the Fermi level is small, which makes Dirac systems comparatively robust
against weak interactions. However, at strong coupling Gross-Neveu
(GN) quantum criticality~\citep{Gross1974,ZinnJustin1991} has received
significant attention~\citep{Herbut2006,Herbut2009, Herbut2009b, Jurivcic2009, Weeks2010,Semenoff2012,Assaad2013,Han2018,Ihrig2018,Lang2019,Parthenios2023,Biedermann2025,Hawashin2025,Huang2025,Lang2025}. At the GN critical point, fermions spontaneously acquire a mass as a consequence of spontaneous symmetry breaking.
In single-layer graphene, sufficiently
strong Coulomb interaction was expected to break the sub-lattice symmetry
that protects the Dirac point~\citep{Herbut2006,Herbut2009, Herbut2009b, Jurivcic2009,Weeks2010, Semenoff2012}.
The magnitude of the fine-structure constant in this system does not, however, appear to be sufficiently large to reach the GN critical point.
The situation is more promising in twisted two-dimensional materials, such as twisted double-layer WSe$_{2}$~\cite{Ma2024}, which is considered an effective strongly-correlated version of single-layer graphene~\cite{Angeli2021,Pan2023}, or in twisted
bilayer graphene (TBG)~\citep{Hawashin2025,Huang2025}. 
In both systems one expects a spontaneous mass generation at neutrality as
function of the twist angle. The nature of the insulating state in  WSe$_2$ is  unclear with some evidence favoring a magnetic state~\cite{Ma2024}. In twisted bilayer graphene there are arguments in favor of an inter-valley-coherent insulator~\cite{Bultinck2020,Liu2021,Ledwith2021strong} that was  observed in experiment, albeit at a different filling~\cite{Nuckolls2023}.

\begin{figure}
    \centering
    \includegraphics[width=0.95\linewidth]{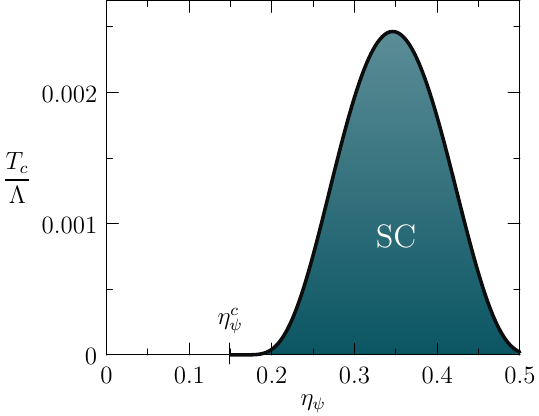}
    \caption{Superconducting transition temperature $T_c$ as function of the anomalous dimension $\eta_\psi$ of the fermions at the Gross-Neveu
mass-generating transition of two-dimensional Dirac systems within
the generalized SYK  approach of this paper. For system
with  $\eta_\psi$ larger than a threshold value $\eta_\psi^c$, fluctuations of the mass-generating boson give rise to a superconducting instability,
where the largest $T_c$  occurs at intermediate values of $\eta_\psi$. $\Lambda$ is the upper cut off of the Dirac theory.}
    \label{fig:TcCurve}
\end{figure}


The mass generation at a GN-critical point is accompanied by strong
quantum fluctuations that give rise to anomalous dimensions $\eta_{\psi}$
and $\eta_{\phi}$ of the fermions and mass-generating collective
bosons, respectively. They are defined through the infrared momentum
dependence of the fermionic ($G\left(k\right)$) and bosonic ($D\left(k\right)$)
propagators:
\begin{eqnarray}
G\left(k_{\mu}\right) & \propto & -i\frac{k_{\mu}\gamma^{\mu}}{\left|k\right|^{2-\eta_{\psi}}},\label{eq:propagators}\\
D\left(k_{\mu}\right) & \propto & \frac{1}{\left|k\right|^{2-\eta_{\phi}}}.\label{eq:GN-bose}
\end{eqnarray}
$k_{\mu}=\left(\omega/v_F,k_{x},k_{y}\right)$ is the (Euclidean) three-vector
of magnitude $\left|k\right|$ and $\gamma^{\mu}$ are the $4\times4$
Dirac matrices. When $\eta_{\psi}$ is finite (and not parametrically
small) the critical fermions, described by Eq.~\eqref{eq:propagators},
are no longer a well-defined quasiparticles. The potential implications
of this break down of the quasiparticle picture and the possibility
of superconductivity masking the GN-critical point are the main
foci of this paper.

Analytic approaches to determine $\eta_{\psi}$ and $\eta_{\phi}$
at GN-criticality are usually based on expansions in small $\epsilon=3-d$
or $1/N$, where $N$ is the number of fermion flavors~\citep{ZinnJustin1991}. Specifically,
one obtains to leading order in $\epsilon$ or $1/N$  the
results $\eta_{\psi}\sim {\cal O}(\epsilon) \sim {\cal O}(1/N)$ and $\eta_{\phi}\sim {\cal O}(\epsilon) \sim 1-{\cal O}(1/N)$~\citep{ZinnJustin1991,Ihrig2018}.
Hence, in the limit of many
fermion flavors, the critical boson behaves highly non-Gaussian while fermionic Dirac particles at criticality are almost well defined
fermionic excitations. Both are well-defined near three space dimensions.  Monte-Carlo simulations for $d=2$~\cite{Chandrasekharan2013,Lang2019,Lang2025}, conformal bootstrap approaches~\cite{Iliesiu2018}, high-loop
expansion in $\epsilon$~\cite{Ihrig2018}, as well as functional renormalization group
approaches~\cite{Vacca2015,Knorr2016}, all yield  values for the anomalous fermion dimension that
range, depending on the problem under consideration up to $\eta_{\psi}\sim 0.17$.
Hence, in two dimensions fermions at the GN-quantum critical point
are in fact ill-defined quasiparticles. 

Recently, an analytic strong-coupling approach to Dirac systems was
proposed in Ref.~\citep{Kim2021} as a generalization of a zero-dimensional
Yukawa-Sachdev-Ye-Kitaev model~\citep{sachdev1993,georges2000,sachdev2010,kitaev2015,Esterlis2019,wang2020,Hauck2020,Esterlis2025}. 
The system displays quantum
chaos~\citep{Kim2021} akin to what was observed in black hole problems~\cite{Maldacena2016}.
For closely related strong-coupling approaches to compressible Fermi systems, see Ref.~\cite{Chowdhury2018,Chowdhury2020,Esterlis2021,Patel2023,Li2024,Esterlis2025}. 
The model of  Ref.~\citep{Kim2021} contains
$M$ real scalar boson fields and $N$ four-component Dirac fermions,
interacting via a Yukawa interaction. Crucially, the couplings of
this interaction are Gaussian random numbers that are the same everywhere
in space, i.e. translation invariance is not broken, even for a specific
realization of the coupling constants. 
Analyzing the averaged model then allows for large $M$ and $N$ to
obtain analytic control of a strong coupling fixed point with a fermionic
anomalous dimension that is not parametrically small. 
The detailed
properties of the fixed point depend on the ratio $ M/N$,
where the behavior near $M\ll N$ behaves similar to the above large-$N$
limit with $\eta_{\psi}\sim{\cal O}\left(M/N\right)$ and $\eta_{\phi}\sim 1- {\cal O}\left(M/N\right)$,
while $M\gg N$ yields $\eta_{\psi}\sim \tfrac{1}{2}-{\cal O}\left(N/M\right)$
and $\eta_{\phi}\sim{\cal O}\left(N/M\right)$. Hence,  within this approach, fermions indeed possess a sizable anomalous dimension $0<\eta_{\psi}<\frac{1}{2}$ and are no-longer well defined quasiparticles.  Ref.~\citep{Kim2021} also demonstrated quantitative agreement of the anomalous exponents obtained from the generalized SYK approach with results obtained using conformal bootstrap~\cite{Iliesiu2018}.

In this paper we generalize the approach of Ref.~\citep{Kim2021} to include superconductivity and analyze
whether strong quantum fluctuations and non-quasiparticle behavior
of the fermions give rise to superconductivity as a secondary instability.
Hence, we examine the intriguing possibility that such a system, devoid
of carriers at zero temperature, might
nevertheless exhibit superconductivity. Interestingly we find that superconductivity occurs provided that the anomalous fermion dimension exceeds a threshold value which in our theory turns out to be 
\begin{equation}
    \eta_{\psi}^c \approx 0.14628.
    \label{eq:eta_crit}
\end{equation} 
Hence, we obtain the curious result that well-defined quasiparticles do not superconduct while ill-defined ones do. Whether the precise value of $\eta_{\psi}^c$ is indeed the correct one to compare with numerical work on GN-models is unclear, given that we are really only solving an approximate strong-coupling theory. What is more important is the very existence of such a threshold value for superconductivity that is comparable to values in two-dimensional systems.
The resulting behavior  for $\eta_\psi>\eta_\psi^c$ is illustrated in Fig.~\ref{fig:TcCurve}, where we show the superconducting transition temperature as function of $\eta_\psi$. 
We expect superconductivity only near the critical GN transition $g=g_c$, i.e.  expect a superconducting dome near $g_c$. Below $g_c$ superconductivity must disappear at some threshold value~\cite{Kopnin2008,Abah2011}, while $T_c$ should vanish exponentially in the gapped phase above $g_c$. 
Hence, we expect that GN criticality should be accompanied by pairing or, at the very least, by enhanced pairing fluctuations. Moreover, our analysis shows how to identify the critical bosonic modes most likely to give rise to an unconventional pairing state.
It is an interesting question to explore whether the superconductivity identified here is related to the seeds of pairing in the ordered state of some Dirac systems where skyrmion configurations of the order parameter were shown to be charge-$2e$ degrees of freedom~\cite{Wiegmann1999,Abanov2000theta,Grover2008,Christos2020,Khalaf2021charged,Ledwith2021strong}.  

We explore generic four-component Dirac spinors, systematically analyse the full set of interactions that can drive the system into a quantum critical regime, identify the leading pairing instabilities that emerge, and discuss the role of phase fluctuations. To facilitate the identification of possible superconducting states, we present generic algebraic conditions on the spinor structure of the pairing state, formulated for an arbitrary Dirac theory with a general number of spinor components coupled to a fully general set of critical bosons. In this context we demonstrate that no pairing arises in the simplest $2+1$-dimensional Dirac theory with a single two-component spinor.
Broadly speaking, the more intricate the Dirac structure, the more readily pairs can form in a manner consistent with Fermi statistics. This point will be made particularly clear in a separate note~\cite{Stangier2025}, which investigates the pairing states of Dirac models for twisted double-layer WSe$_{2}$
and twisted bilayer graphene near Gross-Neveu criticality.

\section{Model and large-$N,M$ saddle point}

\subsection{The model, symmetries, and Dirac spinor structure}

Following Ref.~\citep{Kim2021}, we consider the Hamiltonian of $N$ massless Dirac electrons, each
described by a four component Dirac spinor $\psi_{l}\left(\boldsymbol{x}\right)$
and additional $N$ flavors, i.e. $l=1\cdots N$, that interact with
$M$ massive scalar bosonic modes $\phi_{s}\left(\boldsymbol{x}\right)$
with $s=1\cdots M$. The Hamiltonian of the problem is then given
as
\begin{eqnarray}
H & = & v_{{\rm F}}\sum_{l=1}^{N}\int d^{2}x\psi_{l}^{\dagger}\left(\boldsymbol{x}\right)\left(-i\nabla\right)\cdot\boldsymbol{\alpha}\psi_{l}\left(\boldsymbol{x}\right)\nonumber \\
 & + & \frac{1}{2}\sum_{s=1}^{M}\int d^{2}x\left(\pi_{s}^{2}\left(\boldsymbol{x}\right)+\omega_{0}^{2}\phi_{s}^{2}\left(\boldsymbol{x}\right)+v_{{\rm B}}^{2}\left(\nabla\phi_{s}\left(\boldsymbol{x}\right)\right)^{2}\right)\nonumber \\
 & + & \frac{1}{N}\sum_{lms}\int d^{2}x\left(g_{lm,s}\psi_{l}^{\dagger}\left(\boldsymbol{x}\right)\Upsilon \psi_{m}\left(\boldsymbol{x}\right)\phi_{s}\left(\boldsymbol{x}\right)+h.c.\right),
\end{eqnarray}
where $\psi_{l}\left(\boldsymbol{x}\right)$ is a four-component
Dirac spinor with the additional flavor index $l=1\cdots N$. $\pi_{s}$ is the conjugated momentum to the boson field $\phi_{s}$, i.e. $[\phi_{s}\left(\boldsymbol{x}\right),\pi_{s'}\left(\boldsymbol{x'}\right)]=i\delta_{s,s'}\delta(\boldsymbol{x}-\boldsymbol{x}')$. $v_{{\rm F}}$
and $v_{{\rm B}}$ are the fermionic and bosonic velocities. 
For the
$4\times4$ Dirac matrices hold the usual properties $\left\{ \alpha_{i},\alpha_{j}\right\} =2\delta_{ij}$, $\left\{ \alpha_{i},\beta\right\} =0$, and $\beta^{2}=1$. Curly
brackets stand for the anti-commutator. In our analysis we will also
use the Euclidean covariant formulation with $\gamma^{0}=\beta$ and
$\gamma^{i}=-i\beta\alpha_{i}$. They obey $\left\{ \gamma^{\mu},\gamma^{\nu}\right\} =2\delta^{\mu\nu}$. In what follows we set $v_{\rm F}=1$ while the magnitude of the boson velocity will not enter the universal low-energy behavior. Within the approach of this paper, the additional inclusion of a nonlinear, $\phi^4$, term in the Hamiltonian will not change the universal behavior and is therefore ignored at the outset.

The fermion-boson coupling is determined by the  $4\times 4$ matrix $\Upsilon$. 
We consider four different Hermitian coupling matrices $\Upsilon_i$, $i=1\cdots4$, where boson condensation
induces a gap with the usual spectrum 
\begin{equation}
E=\pm\sqrt{p^{2}+m_{{\rm F}}^{2}}\label{eq:gap_spectrum},
\end{equation}
and fermion mass $m_{{\rm F}}\propto\left\langle \phi\right\rangle $. This is
obviously the case for $\Upsilon_1=\gamma^0$. In total there are four possible interactions that anticommute with $\alpha^{1}$
and $\alpha^{2}$ and hence induce, upon condensation of $\phi$, an isotropic gap at the Dirac
point like in Eq.~\eqref{eq:gap_spectrum}:
\begin{equation}
\left\{\Upsilon_{1},\Upsilon_{2},\Upsilon_{3},\Upsilon_{4}\right\}=\left\{ \gamma^{0},i\gamma^{0}\gamma^{3},i\gamma^{1}\gamma^{2},i\gamma^{0}\gamma^{5}\right\}, \label{eq:coupling_matrices}
\end{equation}
with $\gamma^{5}=\gamma^{0}\gamma^{1}\gamma^{2}\gamma^{3}$.
The transformation of these coupling matrices under parity and time reversal are listed in Table~\ref{tab:pairing_states} below. Notice,
we consider a $2+1$ dimensional system. $\Upsilon_1=\gamma^{0}$ and
$\Upsilon_4=i\gamma^{0}\gamma^{5}$ are allowed coupling matrices even
if one included the third spatial dimension. They correspond to the usual and chiral mass of $3+1$-dimensional Dirac systems~\cite{thaller2013dirac}. $\Upsilon_2=i \gamma^{0},\gamma^{3}$
and $\Upsilon_3=i \gamma^{1},\gamma^{2}$ are additional
coupling terms that only exist in two dimensions. Another distinction of these two pairs of couplings is that $\Upsilon_{1,4}$ anti-commute with $\gamma^5$ (flip  chirality) while $\Upsilon_{2,3}$ commute with with $\gamma^5$ (preserve chirality), i.e. $\Upsilon_{i} \gamma^5=b_i \gamma^5 \Upsilon_i$, with $b_{1,4}=-1$ and $b_{2,3}=1$.

Finally we discuss the random coupling constants $g_{lms}$. Following
Refs.~\citep{Esterlis2019,Hauck2020},  we consider real coupling constants with
\begin{eqnarray}
\overline{g_{lms}g_{l'm's'}} & = & g^{2}\delta_{ss'}\left(\delta_{ll'}\delta_{mm'}+\delta_{lm'}\delta_{ml'}\right).\label{eq:random_g_distr}
\end{eqnarray}
The fact that
the coupling constants are $\boldsymbol{x}$-independent implies that
translation invariance is not broken, even for specific realizations
of the $g_{lms}$. In other words we consider the average of an ensemble
of interacting systems with slightly different values of the coupling
constant. This rather peculiar large-$N$, $M$ limit has the advantage
that it offers  controlled insights into the strong-coupling behavior
of Yukawa coupled theories. 

Before we analyze the Dirac
problem we collect some information of its behavior under symmetry
operations. This will prove  useful for the analysis of the pairing
states. We consider a problem without broken parity and time-reversal symmetry
on the level of the Hamiltonian. Let ${\cal R}$ be a spatial symmetry operation of the problem,
it holds in momentum space:
\begin{eqnarray}
\psi\left(\boldsymbol{k}\right) & \xrightarrow{{\cal R}} & u_{R}\psi\left(R_{v}^{-1}\boldsymbol{k}\right),\nonumber \\
\psi^{\dagger}\left(\boldsymbol{k}\right) & \xrightarrow{{\cal R}} & \psi_{l}^{\dagger}\left(R_{v}^{-1}\boldsymbol{k}\right)u_{R}^{\dagger},\nonumber \\
\phi\left(\boldsymbol{k}\right) & \xrightarrow{{\cal R}} & u_{\phi,R}\phi\left(R_{v}^{-1}\boldsymbol{k}\right),
\end{eqnarray}
with $R_{v}$, $u_{R}$, and $u_{\phi,R}$ the vector, spinor, and
bosonic field representation of ${\cal R}$, respectively. Here we
suppressed the additional flavor indices $l$ and $s$ for simplicity.
For fermionic bilinear combinations 
\begin{equation}
I_{O}=\int d^{d}k\psi^{\dagger}\left(\boldsymbol{k}\right)O\left(\boldsymbol{k}\right)\psi_{l}\left(\boldsymbol{k}\right),
\end{equation}
it holds that the single particle operator $O$, which is a $4\times4$
matrix in spinor space, transforms like $O\left(\boldsymbol{k}\right)\xrightarrow{{\cal R}}u_{R}^{\dagger}O\left(R_{v}\boldsymbol{k}\right)u_{R}$.
Parity transformation is given as
\begin{eqnarray}
\psi\left(\boldsymbol{k}\right) & \xrightarrow{{\cal P}} & u_{P}\psi\left(-\boldsymbol{k}\right),\nonumber \\
\psi^{\dagger}\left(\boldsymbol{k}\right) & \xrightarrow{{\cal P}} & \psi^{\dagger}\left(-\boldsymbol{k}\right)u_{P}^{\dagger},\nonumber \\
\phi\left(\boldsymbol{k}\right) & \xrightarrow{{\cal P}} & p_{i}\phi\left(-\boldsymbol{k}\right),
\end{eqnarray}
with $u_{P}=\beta=\gamma^{0}$. $p_{i}=\pm1$ is the boson parity that depends on the choice of the coupling matrix $\Upsilon_i$.
For  $\Upsilon_1=\beta$ the boson is of even parity.
More generally follows that $u_{P}\Upsilon_i u_{P}^{\dagger}=p_{i}\Upsilon_i$.

Time reversal, ${\cal T}={\cal K}U_{T}^{\dagger}$ is the combination
of complex conjugation ${\cal K}$ and a unitary operator $U_{T}$.
In our model it corresponds for the fermion and boson operators
to
\begin{eqnarray}
\psi\left(\boldsymbol{k},t\right) & \xrightarrow{{\cal {\cal T}}} & u_{T}\psi\left(-\boldsymbol{k},-t\right),\nonumber \\
\psi^{\dagger}\left(\boldsymbol{k},t\right) & \xrightarrow{{\cal {\cal T}}} & \psi^{\dagger}\left(-\boldsymbol{k},-t\right)u_{T}^{\dagger},\nonumber \\
\phi\left(\boldsymbol{k},t\right) & \xrightarrow{{\cal {\cal T}}} & \tau_{i}\phi\left(-\boldsymbol{k},-t\right),
\end{eqnarray}
with $\tau_{i}=\pm1$ the parity of the boson under time reversal, which  depends on the choice of $\Upsilon_i$.
$u_{T}$ is the spinor representation of $U_{T}$ and the operators
are given in the Heisenberg picture. To ensure that the Hamiltonian
is time reversal symmetric it must hold 
\begin{eqnarray}
u_{T}^{\dagger}\boldsymbol{\alpha}^{*}u_{T} & = & -\boldsymbol{\alpha},\nonumber \\
u_{T}^{\dagger}\beta^{*}u_{T} & = & \beta, \nonumber \\
u_{T}^{\dagger}\Upsilon_i^{*}u_{T} & = & \tau_{i}\Upsilon_i.\label{eq:TR_cond}
\end{eqnarray}
The specific representation of $u_{T}$ depends on the choice of the
Dirac matrices. Since time reversal for the fermions squares to ${\cal T}^{2}=-1$
it holds $u_{T}^{*}u_{T}=-1$ and, using $u_{T}^{\dagger}u_{T}=1$,
it follows $u_{T}^{T}=\left(u_{T}^{-1}\right)^{*}=-u_{T}$. Time-reversal
is not a spatial symmetry and must, hence, commute with all operations
of the symmetry group, $\left[{\cal R},{\cal T}\right]=0$. This implies
for the spinor representation$\left[u_{T}^{\dagger}{\cal K},u_{R}\right]=0$
for all symmetry operations ${\cal R}$, and leads to:
\begin{equation}
u_{T}^{\dagger}u_{R}^{*}=u_{R}u_{T}^{\dagger}.\label{eq:time and space}
\end{equation}
a relation that will prove handy as we describe pairing instabilities.

In the context of superconductivity, one encounters anomalous bilinear
forms of the kind 
\begin{equation}
I_{\Delta}=\int d^{d}k\psi^{\dagger}\left(\boldsymbol{k}\right)\Delta\left(\boldsymbol{k}\right)\left(\psi^{\dagger}\left(-\boldsymbol{k}\right)\right)^{T}
\end{equation}
or its Hermitian conjugate. Because of the two creation operators,
it now follows that under symmetry transformations $\Delta\left(\boldsymbol{k}\right)\xrightarrow{{\cal R}}u_{R}^{\dagger}\Delta\left(R_{v}\boldsymbol{k}\right)u_{R}^{*}.$
This is distinct from the behavior of a usual fermion bilinear and not
very convenient. It can be avoided if we use, instead of $\Delta$,
the quantity
\begin{equation}
\Phi\left(\boldsymbol{k}\right)=\Delta\left(\boldsymbol{k}\right)u_{T}^{\dagger}.\label{eq:Delta to Phi}
\end{equation}
Using Eq.~\eqref{eq:time and space}, which relates spatial symmetry
operations and time reversal, one easily finds that $\Phi$ transforms
like a usual bilinear form
\begin{equation}
\Phi\left(\boldsymbol{k}\right)\xrightarrow{{\cal R}}u_{R}^{\dagger}\Phi\left(R_{v}\boldsymbol{k}\right)u_{R}.
\end{equation}
It is this property that makes the Nambu spinor 
\begin{equation}
\Psi\left(\boldsymbol{k}\right)=\left(\begin{array}{c}
\psi\left(\boldsymbol{k}\right)\\
u_{T}^{\dagger}\left(\psi^{\dagger}\left(-\boldsymbol{k}\right)\right)^{T}
\end{array}\right)\label{eq:Nambu spinor}
\end{equation}
the natural, symmetry-adapted description of a superconductor.

For the subsequent analysis of pairing states, it will prove convenient to expand generic Hermitian $4\times4$ matrices in terms
of  $16$ base matrices; see e.g. Refs.~\citep{thaller2013dirac,Nieves2004}:
\begin{eqnarray}
\Gamma_{S} & = & 1,\nonumber \\
\Gamma_{V}^{\mu} & = & \gamma^{\mu},\nonumber \\
\Gamma_{T}^{\mu\nu} & = & \sigma^{\mu\nu}\,\,\,\mu<\nu,\nonumber \\
\Gamma_{A}^{\mu} & = & i\gamma^{\mu}\gamma_{5},\nonumber \\
\Gamma_{P} & = & \gamma_{5},\label{eq:base matrices}
\end{eqnarray}
with $\sigma^{\mu\nu}=\frac{i}{2}\left[\gamma^{\mu},\gamma^{\nu}\right]$. $S$,
$V$, $T$, $A$, and $P$ stand for scalar, vector, (antisymmetric)
tensor, axial vector, and pseudo scalar, respectively. It holds ${\rm \frac{1}{4}tr}\left(\Gamma_{I}^{r}\Gamma_{J}^{r'}\right)=\delta_{I,J}\delta_{r,r'}$
as well as $\left(\Gamma_{J}^{r}\right)^{2}=1$. The coupling matrices of Eq.~\eqref{eq:coupling_matrices}
are $\Upsilon_i\in\left\{ \Gamma_{V}^{0},\Gamma_{T}^{03},\Gamma_{T}^{12},\Gamma_{A}^{0}\right\} $.
Under time reversal, $\Gamma_{S}$, $\Gamma_{V}^{\mu}$, and
$\Gamma_{P}$ are even, while $\Gamma_{T}^{\mu\nu}$ and $\Gamma_{A}^{\mu}$
are odd . In addition $\Gamma_{S}$, $\Gamma_{V}^{0},\Gamma_{T}^{ij}$
and $\Gamma_{A}^{i}$ are even under parity while $\Gamma_{V}^{i}$,
$\Gamma_{T}^{0i}$, $\Gamma_{A}^{0}$, and $\Gamma_{P}$ are odd under
parity. Here, the superscript  $i$ stands for the spatial components.

\subsection{Large-$N$, $M$ saddle point equations}

The analysis of the large-$N$ equations is a direct extension of
the approach of Refs.~\citep{Esterlis2019,wang2020,Esterlis2025}. We perform the
replica trick to average over the random variables $g_{lms}$,  drawn from a Gaussian ensemble governed by Eq.~\eqref{eq:random_g_distr},
and take the replica-symmetric limit. We then introduce the bi-local
fields ($G$ and $F$ have a $4\times4$ matrix structure)
\begin{eqnarray}
G\left(x,x'\right) & = & \frac{1}{N}\sum_{l}\psi_{l}\left(x\right)\odot\psi_{l}^{\dagger}\left(x'\right),\nonumber \\
F\left(x,x'\right) & = & \frac{1}{N}\sum_{lc}\psi_{l}\left(x\right)\odot\left(\psi_{l}\left(x'\right)u_{T}\right),\nonumber \\
D\left(x,x'\right) & = & \frac{1}{M}\sum_{s}\phi_{s}\left(x\right)\phi_{s}\left(x'\right),\label{eq:bilocal}
\end{eqnarray}
that describe normal and pairing correlations of the fermions as well
as bosonic correlations. $x=\left(\tau,\boldsymbol{x}\right)$ combines
imaginary time and spatial coordinates. In our definition of the anomalous
function $F$ we added $u_{T}$, following the discussion that led
to Eq.~\eqref{eq:Delta to Phi}.

Integrating out the primary fields gives rise to the saddle point equations in the limit of large
$N$ and $M$ at fixed $N/M$. As we have to include pairing terms
in the analysis, see Ref.~\citep{Esterlis2019}, we use the Nambu spinor
of Eq.~\eqref{eq:Nambu spinor}. At the saddle point it holds 
\begin{eqnarray}
\Sigma\left(x\right) & = & g^{2}\frac{M}{N}\Upsilon_i G\left(x\right)\Upsilon_i D\left(x\right),\label{eq:normalself energy}\\
\Phi\left(x\right) & = & -\tau_{i}g^{2}\frac{M}{N}\Upsilon_i F\left(x\right)\Upsilon_i D\left(x\right),\label{eq:anomalous self energy}\\
\Pi\left(x\right) & = & -g^{2}{\rm tr}\left(\Upsilon_i G\left(x\right)\Upsilon_i G\left(-x\right)\right)\nonumber \\
  & + & \tau_{i}g^{2}{\rm tr}\left(\Upsilon_i\bar{F}\left(x\right)\Upsilon_i F\left(-x\right)\right).
 \label{eq:boson self energy}
\end{eqnarray}
$\tau_{i}$ is the transformation of the
boson $\phi$, and hence of the coupling matrix $\Upsilon_i$, under time
reversal; see Eq.~\eqref{eq:TR_cond}. $\Sigma$, $\Phi$, and $\Pi$
formally enter the theory as Lagrange-parameter fields that enforce
Eq.~\eqref{eq:bilocal}. At the saddle point they play the role of
the fermionic and bosonic self energies and are determined by the
Dyson equations which we write in the frequency and momentum domain
with $k=\left(\omega,\boldsymbol{k}\right)$ as:
\begin{eqnarray}
\hat{G}^{-1}\left(k\right) & = & \hat{G}_{0}^{-1}\left(k\right)-\hat{\Sigma}\left(k\right),
\end{eqnarray}
where
\begin{equation}
\hat{G}\left(k\right)=\left(\begin{array}{cc}
G\left(k\right) & F\left(k\right)\\
\bar{F}\left(k\right) & -u_{T}^{\dagger}G^{T}\left(-k\right)u_{T}
\end{array}\right),
\end{equation}
and similar structure for $\hat{\Sigma}$ in terms of $\Sigma$ and
$\Phi$. The inverse of the bare fermionic propagator in Nambu space is 
\begin{equation}
  \boldsymbol{G}_{0}^{-1}\left(k\right)=\left(\begin{array}{cc}
ik_{\mu}\gamma^{\mu}\gamma^{0} & 0\\
0 & ik_{\mu}\gamma^{0}\gamma^{\mu}
\end{array}\right)
\end{equation}
The bosonic Dyson equation is given as
\begin{equation}
D\left(q\right)=\frac{1}{\omega_{0}^{2}+q^{2}-\Pi\left(q\right)}.
\end{equation}

\subsection{Normal state analysis}

We first analyze the solution of the saddle-point equations in the normal
state where $\Phi=F=0$. Except for the slight generalization on the coupling
matrix $\Upsilon_i$ this analysis was performed in Ref.~\citep{Kim2021}.
We perform the analysis at $T=0$ and assume that the normal state
is critical, i.e. we analyze the problem right at the GN-critical point, which corresponds to a critical value of the coupling constant $g_c \sim \omega_0\Lambda^{-1/2}$ with upper momentum cut off $\Lambda$.
In our analysis of  the critical state the following
expressions for Fourier transformations will be useful:
\begin{eqnarray}
\int d^{3}x\left|x\right|^{-2a}e^{ix_{\mu}q^{\mu}} & = & C_{a}\left|q\right|^{2a-3},\nonumber \\
\int d^{3}xx_{\mu}\gamma^{\mu}\left|x\right|^{-2a-1}e^{ix_{\mu}q^{\mu}} & = & -iB_{a}q_{\mu}\gamma^{\mu}\left|q\right|^{2a-4},
\end{eqnarray}
where $C_{a}=\frac{2^{3-2a}\pi^{3/2}\Gamma\left(\frac{3}{2}-a\right)}{\Gamma\left(a\right)}$
and $B_{a}=\frac{C_{a-\frac{1}{2}}}{1-2a}$. The analogous expressions of the
inverse transforms follow immediately.

\begin{figure}
    \centering
\includegraphics[width=0.8\linewidth]{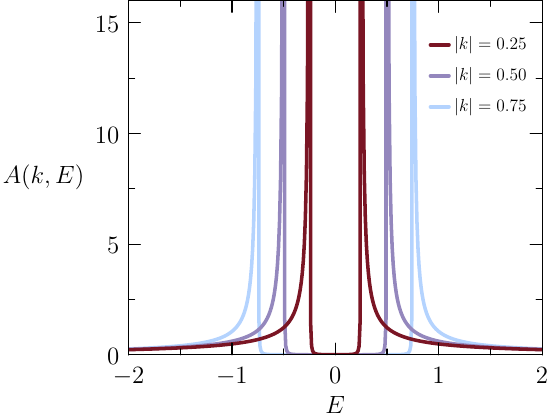}
    \caption{Normal state single-particle spectral function $A(\boldsymbol{k}, E) = -{\rm Im}\big[ {\rm Tr}G(\boldsymbol{k}, i\omega\to E+i0^+)]$  vs energy $E$, showing a strong broadening due to the self energy Eq.~(\ref{eq:fullselfenergy}) that
    indicates the absence of well-defined quasiparticles. 
    For this plot we chose amplitude $A = 1$, exponent $\eta_\psi = 0.14628$, and wave-vector $|\boldsymbol{k}|= 0.25$ (red), $|\boldsymbol{k}| = 0.50$ (purple),    and $|\boldsymbol{k}|=0.75$ (blue).  
    }
    \label{fig:normalspectral}
\end{figure}

\begin{figure}
    \centering
    \includegraphics[width=0.8\linewidth]{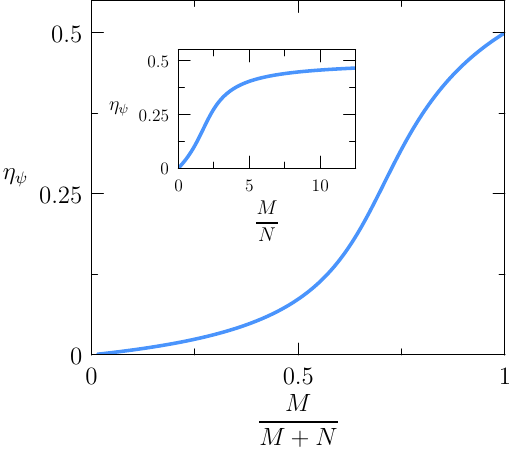}
    \caption{Dependence of the anomalous exponent of the fermion spectrum on the ratio $M/N$ of the bosonic and fermionic flavors, respectively (notice, as we consider four-component Dirac spinors, we have in total $4N$ fermion flavors). Crucially, the approach used here allows for a controlled analysis of $\eta_\psi$ that is not parametrically small and reaches values up to $\eta_\psi=\tfrac{1}{2}$. }
    \label{fig:eta_over_MoN}
\end{figure}

For a critical system we make the power law ansatz  
\begin{eqnarray}
\label{eq:fullselfenergy}
\Sigma\left(k\right) & = & -A\left(i\omega-v\boldsymbol{k}\cdot\boldsymbol{\alpha}\right)\left|k\right|^{-\eta_\psi},\nonumber \\
 & = & -iAk_{\mu}\gamma^{\mu}\gamma^{0}\left|k\right|^{-\eta_\psi},
 \label{eq:SEansatz}
\end{eqnarray}
with amplitude $A>0$ and  exponent $\eta_\psi$. We determine the exponent from
the coupled saddle point equations. For $\eta_\psi>0$ the self energy
has a frequency and momentum dependence that strongly
broadens quasiparticle excitations, 
as seen in Fig.~\ref{fig:normalspectral} for the 
normal state spectral function.
In the infrared regime $\Sigma\left(k\right)$ 
dominates over the bare propagator and we obtain from the Dyson equation $G\left(k\right)^{-1}\approx-\Sigma\left(k\right)$ which
yields
\begin{equation}
G\left(k\right)=-\frac{i}{A}\gamma^{0}\frac{k_{\mu}\gamma^{\mu}}{\left|k\right|^{2-\eta_\psi}}.\label{eq:IR-1}
\end{equation}
Fourier transformation to coordinate space yields the result
\begin{equation}
    G\left(x\right)=-\frac{B_{\tfrac{1}{2}(1-\eta_\psi)}}{\left(2\pi\right)^{3}A}\gamma^{0}\frac{\gamma^{\mu}x_{\mu}}{\left|x\right|^{3+\eta_\psi}}.
\end{equation}
For the bosonic propagator follows from Eq.~\eqref{eq:boson self energy}
\begin{eqnarray}
\Pi\left(x\right) & = & 4g^{2}\frac{B_{\frac{1-\eta_\psi}{2}}^{2}}{\left(2\pi\right)^{6}A^{2}}\frac{c_{\mu\nu}x_{\mu}x_{\nu}}{\left|x\right|^{6+2\eta_\psi}},\label{eq:bubble}
\end{eqnarray}
where $c_{\mu\nu}=\frac{1}{4}{\rm tr}\left(\Upsilon_i\gamma^{0}\gamma^{\mu}\Upsilon_i\gamma^{0}\gamma^{\nu}\right)$. For the four coupling matrices $\Upsilon_i$ of Eq.~\eqref{eq:coupling_matrices} that anti-commute with the $\alpha_i$
follows that  $c_{\mu\nu}=\delta_{\mu\nu}$. Returning to momentum
space we separately analyze the zero momentum contribution and the
dynamic part 
\begin{equation}
\Pi\left(q\right)=\Pi\left(0\right)+\delta\Pi\left(q\right).
\end{equation}
Being in a critical state it must hold that the renormalized boson
frequency
\begin{equation}
\omega_{r}^{2}=\omega_{0}^{2}-\Pi\left(0\right),
\end{equation}
vanishes at $T=0$.  We then obtain
\begin{equation}
    \Pi\left(q\right)=\omega_0^2-g^{2}\frac{c_{\Pi}}{A^{2}}\left|q\right|^{1+2\eta_\psi}
    \label{eq:Pi_powerlaw}
\end{equation}
with $c_{\Pi}=-4B_{\frac{1-\eta_\psi}{2}}^{2}C_{2+\eta_\psi}/\left(2\pi\right)^{6}$,
which yields for the bosonic propagator
\begin{equation}
D\left(q\right)=\frac{A^{2}}{g^{2}c_{\Pi}}\frac{1}{\left|q\right|^{1+2\eta_\psi}}.
\end{equation}
Note that compared to the GN critical point in Eq.~\eqref{eq:GN-bose} the anomalous exponents of the bosonic and fermionic propagators are connected by $\eta_\phi=1-2\eta_\psi$.
The Fourier transform is $D\left(x\right)=\frac{A^{2}}{g^{2}}c_{D}\left|x\right|^{2\eta_\psi-2}$
with $c_{D}=-\frac{\left(2\pi\right)^{3}C_{\frac{1}{2}+\eta_\psi}}{4B_{\frac{1-\eta_\psi}{2}}^{2}C_{2+\eta_\psi}}$.
This yields for the self energy:
\begin{equation}
\Sigma\left(x\right)=-\frac{AM}{4N}\frac{C_{\frac{1}{2}+\eta_\psi}}{B_{\frac{1-\eta_\psi}{2}}C_{2+\eta_\psi}}\left|x\right|^{\eta_\psi-5}\Upsilon_i\gamma^{0}\gamma^{\mu}x_{\mu}\Upsilon_i
\end{equation}
which we Fourier transform to momentum space and finally obtain
\begin{equation}
    \Sigma\left(k\right)=i\frac{AM}{4N}\frac{C_{\frac{1}{2}+\eta_\psi}B_{\frac{4+\eta_\psi}{2}}}{B_{\frac{1-\eta_\psi}{2}}C_{2+\eta_\psi}}\left|k\right|^{-\eta_\psi}k_{\mu}\Upsilon_i\gamma^{0}\gamma^{\mu}\Upsilon_i.
\end{equation}
This result must be equal to our original ansatz, Eq.~\eqref{eq:SEansatz}, to ensure that it is a self-consistent solution. This is indeed the case if
\begin{equation}
\frac{M}{N}=-4\frac{B_{\frac{1-\eta_\psi}{2}}C_{2+\eta_\psi}}{C_{\frac{1}{2}+\eta_\psi}B_{\frac{4+\eta_\psi}{2}}}=2\frac{ \left(\eta_\psi -3\right) \left(\eta_\psi -1\right) \sin ^2\left(\frac{\pi  \eta_\psi }{2}\right)}{\eta_\psi  \left(2 \eta_\psi +1\right) \cos \left(\pi  \eta_\psi \right)}\label{eq:cond_Delta}
\end{equation}
as well as
\begin{equation}
\gamma^{\mu}=\Upsilon_i\gamma^{0}\gamma^{\mu}\Upsilon_i\gamma^{0}\label{eq:cond_gamma}
\end{equation}
for $\mu=0,1,2$. The first equation determines the exponent $\eta_\psi$. Analyzing the restrictions under which all Fourier integrals are convergent
yields $0<\eta_\psi<\tfrac{1}{2}$ for the exponent.
In Fig.~\ref{fig:eta_over_MoN} we show $\eta_\psi$ as function of the ratio $M/N$. It holds
that $\eta_\psi\rightarrow0$ as $M/N$ goes to zero and $\eta_\psi\rightarrow1/2$
in the opposite limit, $M/N\rightarrow\infty$. For $M=N$ follows
$\eta_\psi\approx0.08658$. As for the second condition, Eq.~\eqref{eq:cond_gamma},
on the coupling matrices, one  easily finds that it is obeyed precisely
for the four $\Upsilon_i$ listed in Eq.~\eqref{eq:coupling_matrices}.
Hence, Eq.~\eqref{eq:SEansatz} is a consistent solution for the normal state energy with $\eta_\psi$ given by Eq.~\eqref{eq:cond_Delta}.

\section{Superconductivity  at the Gross-Neveu critical point}

\subsection{Linearized gap equation}

In order to identify the leading superconducting instabilities we
solve the saddle point Eq.~\eqref{eq:anomalous self energy} with the
anomalous propagator evaluated to linear order in $\Phi$:
\begin{eqnarray}
F\left(p\right) & = & -G\left(p\right)\Phi\left(p\right)u_{T}^{\dagger}G^{T}\left(-p\right)u_{T},\nonumber \\
 & = & -\frac{p_{\mu}p_{\nu}}{A^{2}\left|p\right|^{4-2\eta_\psi}}\gamma^{0}\gamma^{\mu}\Phi\left(p\right)\gamma^{\nu}\gamma^{0}.
\end{eqnarray}
In the last step we used the critical normal state solution for $G\left(p\right)$, discussed
in the previous section. If we furthermore use the corresponding power-law
result for the boson propagator, the linearized version of Eq.~\eqref{eq:anomalous self energy}
becomes
\begin{equation}
\Phi\left(k\right)=\lambda_{{\rm p}}\tau_{i}\int\frac{d^{3}p}{4\pi }\frac{p_{\mu}p_{\nu}\Upsilon_i\gamma^{0}\gamma^{\mu}\Phi\left(p\right)\gamma^{\nu}\gamma^{0}\Upsilon_i}{\left|p\right|^{4-2\eta_\psi}\left|k-p\right|^{1+2\eta_\psi}},\label{eq:lin_gap_f}
\end{equation}
where we introduced the coupling constant of the pairing problem 
\begin{equation}
\lambda_{{\rm {\rm p}}}(\eta_\psi)=\frac{1}{2 \pi^2 c_{\Pi}}\frac{M}{N }.
\label{eq:pairing_coupling}
\end{equation}
Using Eq.~\eqref{eq:cond_Delta} for $M/N$ and $c_\Pi$ as given in the text below Eq.~\eqref{eq:Pi_powerlaw}, it follows that this coupling constant is fully determined by the anomalous exponent $\eta_\psi$ of the fermions; in particular it is independent of the amplitude $A$ of the fermionic propagator and the  value of the coupling constant $g$. For small $\eta_\psi$ it holds $\lambda_{{\rm p}}\approx 3\eta_{\psi}\left(1-\frac{7}{3}\eta_{\psi}\cdots\right)$ while  $\lambda_{{\rm p}}(\eta_\psi\rightarrow \tfrac{1}{2})\rightarrow \tfrac{5}{3\pi}  $. In Fig.~\ref{fig:lambda_crit} we  show the dependence of $\lambda_{{\rm {\rm p}}}$ on
 $\eta_\psi$ for $0<\eta_\psi< \tfrac{1}{2}$, i.e. the entire regime of relevance.

Eq.~\eqref{eq:lin_gap_f} is the linearized gap equation of pairing
in quantum critical Dirac fluids. It determines the momentum and frequency
dependence as well as the spinor structure of the superconducting
state. Both are closely entangled, a property that will be crucial
in our subsequent solution of this integral equation. 
While formally
our analysis is performed at $T=0$, finite temperatures can easily
be reintroduced as one converts the frequency integration to a Matsubara
summation, or, approximately, by introducing a lower cut off $\sim T$
of the frequency integral.

In order to analyze the pairing state we first focus on its structure
in spinor space. Since $\Phi\left(k\right)$ transforms properly under
symmetry operations, it is natural to expand it in terms of the basis
matrices of Eq.~\eqref{eq:base matrices}
\begin{equation}
\Phi\left(k\right)=\sum_{rJ}\chi_{J}^{r}\left(k\right)\Gamma_{J}^{r},\label{eq:Delta_expansion}
\end{equation}
 such that 
\begin{equation}
\chi_{J}^{r}\left(k\right)=\lambda_{{\rm p}}\tau_{i}\sum_{r'J'}\int\frac{d^{3}p}{4\pi }\frac{L_{JJ'}^{rr'}\left(p\right)\chi_{J'}^{r'}\left(p\right)}{\left|p\right|^{2-2\eta_\psi}\left|k-p\right|^{1+2\eta_\psi}},\label{eq:gap_01}
\end{equation}
with 
\begin{eqnarray}
L_{JJ'}^{rr'}\left(p\right) & = & \frac{p_{\mu}p_{\nu}}{4p^{2}}{\rm tr}\left(\Upsilon_i\gamma^{0}\gamma^{\mu}\Gamma_{J}^{r}\gamma^{\nu}\gamma^{0}\Upsilon_i\Gamma_{J'}^{r'}\right).
\label{eq:matrix_L}
\end{eqnarray}
Here, the summation is over $\mu$ and $\nu$, while no summation over $i$, i.e. over the choice of the coupling matrix $\Upsilon_i$, is implied. 
To keep the number of indices manageable we do not indicate explicitly that 
 $L_{JJ'}^{rr'}\left(p\right)$ depends on $i$.  It holds for all four $\Upsilon_i$ that $L_{JJ'}^{rr'}\left(p\right)$
is diagonal w.r.t. $S$, $V$, $T$, $A$, and $P$ that
stand for scalar, vector, (antisymmetric) tensor, axial vector, and
pseudo scalar, respectively, i.e.
\begin{equation}
L_{JJ'}^{rr'}\left(p\right)=l_{J}^{rr'}\left(p\right)\delta_{J,J'}.
\end{equation}
The eigenvalues $l_{J}$ of the matrices $l_{J}^{rr'}\left(p\right)$
are momentum independent and $\pm1$. The eigenvectors only depend
on the direction of $p=\left(\omega,\boldsymbol{p}\right)$ and not
on its magnitude, which we denote as $P=\left|p\right|$. 

Let us analyze these matrices for the distinct pairing symmetries, i.e. for different $J$. If $J=S$ it holds for all four pairing interactions $\Upsilon_i$
\begin{equation}
 l_{S}=1.    
\end{equation}
If the pairing state is of vector
nature, the pairing matrix $l_{V}^{rr'}\left(p\right)$ is 
a $4\times4$ matrix
\begin{equation}
    l_{V}=p_i \left(\begin{array}{cc}
Q_{\mu\nu} & 0\\
0 & b_i
\end{array}\right),
\label{eq:l_Vmatrix}
\end{equation}
with $3\times3$ block $Q_{\mu\nu}=\frac{2q^{\mu}q^{\nu}}{q^{2}}-\delta^{\mu\nu}$
and  $2+1$-dimensional momenta $q^{\mu}$ as well as a $1\times 1$ block. We recall that $p_i$ is the parity of the boson and  $b_i=\pm 1$ determines whether $\gamma^5$ commutes or anti-commutes with the coupling matrix $\Upsilon_i$. Obviously the result depends on the behavior of the pairing matrix $\Upsilon_i$.  The expression for the $6\times 6$ matrix $l_T$, that describes anti-symmetric tensor matrices, is rather lengthy. As we will see,   we will not need it in our subsequent analysis. For pairing states with axial vector character it holds $ l_{A}=b_i l_{V}$, while for pseudo-scalars it holds
\begin{equation}
 l_{P}=b_i.    
\end{equation}
 These results are summarized in Table~\ref{tab:pairing_states}.

\begin{table}
\begin{tabular}{|c|c|c|c|c|}
\hline 
$\Upsilon_i$ & $\gamma^{0}$ & $i\gamma^{0}\gamma^{3}$ & $i\gamma^{1}\gamma^{2}$ & $i\gamma^{0}\gamma^{5}$\tabularnewline
\hline 
\hline 
$p_{i}$ & $1$ & $-1$ & $1$ & $-1$\tabularnewline
\hline 
$\tau_{i}$ & $1$ & $-1$ & $-1$ & $-1$\tabularnewline
\hline 
$b_i$ & $-1$ & $1$ & $1$ & $-1$\tabularnewline
\hline 
\hline
$l_{S}$ & ${\color{ForestGreen}1}$ & $1$ & $1$ & $1$\tabularnewline
\hline 
$l_{V}$ & $\left(\begin{array}{cc}
Q\\
 & -1
\end{array}\right)$ & $\left(\begin{array}{cc}
-Q\\
 & {\color{ForestGreen}-1}
\end{array}\right)$ & $\left(\begin{array}{cc}
{\color{red}Q}\\
 & 1
\end{array}\right)$ & $\left(\begin{array}{cc}
-Q\\
 & 1
\end{array}\right)$\tabularnewline
\hline 
$l_{A}$ & $\left(\begin{array}{cc}
-Q\\
 & 1
\end{array}\right)$ & $\left(\begin{array}{cc}
-Q\\
 & -1
\end{array}\right)$ & $\left(\begin{array}{cc}
Q\\
 & 1
\end{array}\right)$ & $\left(\begin{array}{cc}
Q\\
 & -1
\end{array}\right)$\tabularnewline
\hline 
$l_{P}$ & $-1$ & $1$ & $1$ & ${\color{ForestGreen}-1}$\tabularnewline
\hline 
\hline 
$\Phi$ & $ 1 $ & $ \gamma^3$ & $-$ & $ \gamma^5 $\tabularnewline
\hline 
\end{tabular}

\caption{Four different coupling interactions $\Upsilon_i$ along with the transformation
of the critical boson $\phi$ under parity ($p_{\phi}$) and time
reversal ($\tau_{\phi}$).
In addition we show the matrix structure
of the pairing interaction $l_{J}$ for $J\in\{S,V,A,P\}$. $Q$ refers to the $3\times3$ block $Q_{\mu\nu}=2q^{\mu}q^{\nu}/q^{2}-\delta^{\mu\nu}$.
In {\color{ForestGreen} green}
we  show the interactions that give rise to pairing. For
$\Upsilon_i=i\gamma^{1}\gamma^{2}$ the anisotropic interaction $Q$, marked in {\color{red} red}  and  due to
the pairing in the $l_{V}$ channel, is the dominant pairing interaction.
However, its pairing strength is always smaller
than the threshold coupling  value $\lambda_{p}^{c}$, i.e. no superconducting state emerges for this interaction and  one might at
best observe enhanced pairing fluctuations. The last row shows the spinor structure of the pairing self energy $\Phi$. }
\label{tab:pairing_states}
\end{table}

\subsection{Pairing symmetry and Pauli principle}

Before we solve the linearized gap equation further, we summarize
the implications of the Pauli principle for the pairing amplitudes
$\chi^r_{J}\left(k\right)$. If we consider a pairing expectation value
$\Delta_{ab}\left(k\right)\sim\left\langle \psi_{a}\left(k\right)\psi_{b}\left(-k\right)\right\rangle $,
fermionic anti-commutation yields $\Delta\left(k\right)=-\Delta^{T}\left(-k\right)$
which corresponds with Eq.~\eqref{eq:Delta to Phi} to
\begin{eqnarray}
\Phi\left(k\right) & = & -u_{T}^{*}\Phi^{T}\left(-k\right)u_{T}.\label{eq:Delta_Pauli}
\end{eqnarray}
If we now insert the expansion of Eq.~\eqref{eq:Delta_expansion} into
Eq.~\eqref{eq:Delta_Pauli} and use the orthogonality of the $\Gamma_{J}^{r}$
it follows
\begin{equation}
\chi_{rJ}\left(k\right)=\tau_{J}\chi_{rJ}\left(-k\right),\label{eq:Pauli_fin}
\end{equation}
where $\tau_{J}$ is the parity of $\Gamma_{J}^{r}$ under time reversal,
i.e. $u_{T}^{\dagger}\Gamma_{J}^{rT}u_{T}=\tau_{J}\Gamma_{J}^{r}$
with $\tau_{J}=\pm1$, independent on $r$. To derive Eq.~\eqref{eq:Pauli_fin} we used $u_{T}^{*}=-u_{T}^{\dagger}$
 valid for ${\cal T}^{2}=-1$. Thus, for pairing states with TR-even
spinor structure, $\chi_{rJ}\left(k\right)$ is an even function of
$k=\left(\omega,\boldsymbol{k}\right)$, while it is odd if the pairing
state has a spinor structure that is odd under time reversal. Since
$\tau_{S}=\tau_{V}=\tau_{P}=1$ and $\tau_{T}=\tau_{A}=-1$,
it follows for the Pauli condition
\begin{eqnarray}
\chi_{S}\left(k\right) & = & \chi_{S}\left(-k\right),\nonumber \\
\chi_{V}\left(k\right) & = & \chi_{V}\left(-k\right),\nonumber \\
\chi_{T}\left(k\right) & = & -\chi_{T}\left(-k\right),\nonumber \\
\chi_{A}\left(k\right) & = & -\chi_{A}\left(-k\right),\nonumber \\
\chi_{P}\left(k\right) & = & \chi_{P}\left(-k\right).\label{eq:Pauli_expansion_cond}
\end{eqnarray}
This is analogous to singlet states being even and triplet states
odd under $k\rightarrow-k$ for single orbital pairing~\cite{Sigrist1991}. Below we find that in an expansion in spherical harmonics with respect to the three dimensional vector $k$, only pairing states with  angular momentum $l=0$ occur. This immediately excludes $\chi_{T}\left(k\right)$ and $\chi_{A}\left(k\right)$ which only contain odd $l$.

\subsection{Gap equation in spherical harmonics}

To solve the linearized gap equation, where the spinor structure and the momentum dependence are strongly coupled, we expand the pairing wave function and interaction in spherical harmonics:
\begin{eqnarray}
\chi_{J}^{r}\left(p\right) & = & \sum_{lm}\varphi_{lm}^{r}\left(P\right)Y_{lm}\left(\Omega_{p}\right),\nonumber \\
l_{J}^{rr'}\left(\Omega_{p}\right) & = & \sum_{l m} \Lambda_{l m}^{rr'}
Y_{l  m} \left(\Omega_{q}\right),\nonumber \\
\frac{1}{\left|k-p\right|^{1+2\eta_\psi}} & = & \sum_{l,m}\frac{4\pi z_{l}\left(P/K\right)Y_{lm}\left(\Omega_{k}\right)Y_{lm}^{*}\left(\Omega_{p}\right)}{\left(K^{2}+P^{2}\right)^{1/2+\eta_\psi}}.\label{eq:SHexp}
\end{eqnarray}
Here we dropped, for simplicity, the index $J$ in $\varphi_{lm}^{r}$
and $\Lambda_{l_{2}m_{2}}^{rr'}$ on the right hand side, and used
\begin{equation}
z_{l}\left(s\right)=\frac{1}{2}\int_{-1}^{1}dxP_{l}\left(x\right)\left(1-\frac{2sx}{1+s^{2}}\right)^{-1/2-\eta_\psi},
\end{equation}
with Legendre polynomial $P_{l}\left(x\right)$. It holds $z_{l}\left(s\right)=z_{l}\left(s^{-1}\right)$.
Below we will also employ the asymptotic behavior 
\begin{equation}
z_{l}\left(s\right)=z_{l}^{0}\left\{ \begin{array}{ccc}
s^{l} & {\rm if} & s\ll1\\
s^{-l} & {\rm if} & s\gg1
\end{array}\right. ,\label{eq:alpha asympt}
\end{equation}
with $z_{l}^{0}=\frac{\prod_{l'=0}^{l-1}\left(2l'+\gamma\right)}{\left(2l+1\right)!!}$, implying $z_{l=0}^{0}=1$.

This expansion then yields for the gap equation
\begin{equation}
\varphi_{lm}^{r}\left(K\right)=\lambda_{{\rm p}}\tau_{i}\sum_{r'l'm'}\int_{0}^{\infty}dP\frac{M_{lm,l'm'}^{rr'}z_{l}\left(P/K\right)}{\left(K^{2}+P^{2}\right)^{1/2+\eta_\psi}P^{-2\eta_\psi}}\varphi_{l'm'}^{r'}\left(P\right).
\end{equation}
where we only have a one-dimensional integration over the magnitude
$P=\sqrt{\omega^{2}+\boldsymbol{p}^{2}}$ left. We introduced 
\begin{equation}
    M_{lm,l'm'}^{rr'}=\sum_{l''m''}\Lambda_{l''m''}^{rr'}T_{l''m'',l'm'}^{lm},
\end{equation}
that describes the angular momentum transferred by the interaction.
Analyzing this transferred momentum,  only  $l''=0$ and $l''=2$
occur  since $l_{J}^{rr'}\left(\Omega_{p}\right)$ only contains constant terms or terms that transform like a quadrupole  $p_\mu p_\nu/p^2 $ in space time, see e.g. $l_V^{r,r'}$ of Eq.~\eqref{eq:l_Vmatrix}. Finally, $T_{l''m'',l'm'}^{lm}$ vanishes unless $\left|l''-l'\right|\leq l\leq l''+l'$
when it is expressed in terms of Clebsch-Gordon coefficients:
\begin{eqnarray}
T_{l''m'',l'm'}^{lm} & = & \sqrt{\frac{\left(2l'+1\right)\left(2l''+1\right)}{4\pi\left(2l+1\right)}}\left\langle l''0l'0\mid l0\right\rangle \nonumber \\
 & \times & \left\langle l''m''l'm'\mid lm\right\rangle .
\end{eqnarray}
Next we  solve this pairing problem for the different pairing states.

\subsubsection{Scalar and Pseudo-scalar, and one-component vector pairing states}

The situation is particularly simple for pairing in the scalar and
pseudo-scalar channel $J=S$ and $P$ or for the one-dimensional block of the vector channel $V$, where the indices $r$, $r'$
only take one value; and hence the matrix (or matrix block) obeys $l_{J}^{rr'}=\pm 1$. In this case it holds 
\begin{equation}
    M_{lm,l'm'}=\delta_{l,l'}\delta_{m,m'}.
\end{equation}
Then follows for the linearized gap equation the integral equation
\begin{equation}
\varphi_{lm}\left(K\right)=l_{J}\tau_{i}\lambda_{{\rm p}}\int_{0}^{\infty}dP\frac{z_{l}\left(P/K\right)\varphi_{lm}\left(P\right)}{\left(K^{2}+P^{2}\right)^{1/2+\eta_\psi}P^{-2\eta_\psi}}.\label{eq:47}
\end{equation}
Notice, this integral equation is formally very similar to the one that occurs in compressible quantum critical systems, often referred to as the "$\gamma$-model"~\cite{abanov2001,abanov2001b,ChubukovJS2005,abanov2020-I,abanov2020-II,wu2020-III,wu2021-IV,wu2021-V,Ojajarvi2024}, with exponent $\gamma=1+2\eta_\psi$ (i.e. $1<\gamma <2$). Notice, in compressible systems the variable $K$ corresponds to the fermionic frequency, while it is the Lorentz-invariant magnitude $K=\sqrt{\omega^{2}+\boldsymbol{k}^{2}}$ in our problem. The common feature of both problems is the highly non-local pairing interaction in $K$.

Nontrivial solutions  of Eq.~\eqref{eq:47} require $\tau_{i}l_{J}=+1$, i.e. for time-reversal even bosons $l_{J}=+1$ matters while eigenvalues $l_{J}=-1$ are
important for time-reversal odd bosons. As one can deduce from Table~\ref{tab:pairing_states}, this is the case for $\Upsilon_1=\gamma^{0}$
if $J=S$,  for $\Upsilon_2=i\gamma^{0}\gamma^{3}$ if $J=V$ as long as we consider
the one-dimensional block $-1$, and for $\Upsilon_4=i\gamma^{0}\gamma^{5}$
with $J=P$. These three states are marked with red color in the table. In what follows we consider those cases and set  $\tau_{i}l_{J}=+1$.

We start our analysis by performing an approximate solution of this
integral equation. Following~\cite{ChubukovJS2005,Hauck2020,Ojajarvi2024,Wang_Chub_2025} we transform the integral equation into a differential equation with appropriate boundary conditions.  This is, as we will see, even quantitatively accurate
for $\eta_\psi$ not too small. In addition, it will provide
us with the intuition to solve the problem more accurately. We split
in Eq.~\eqref{eq:47} the contributions for $P<K$ and $P>K$ and treat
them in the limits $P\ll K$ and $P\gg K$. Using Eq.~\eqref{eq:alpha asympt}
this leads to 
\begin{eqnarray}
\varphi_{lm}\left(K\right)&=&\lambda_{{\rm p}}z_{l}^{0}\left(\int_{T}^{K}\frac{dP\varphi_{lm}\left(P\right)}{K^{l+1+2\eta_\psi}P^{-l-2\eta_\psi}}\right. \nonumber \\ & + & \left.  \int_{K}^{\Lambda}\frac{dPK^{l}\varphi_{lm}\left(P\right)}{P^{1+l}}\right),
\end{eqnarray}
with $z_{l}^{0}$ of Eq.~\eqref{eq:alpha asympt}.
Here we introduced the temperature, $T$, as a lower cut off and added
an upper cut off $\Lambda$ above which the power-law behavior ceases
to be correct. In this form, the equation can easily be transformed
into a differential equation. First we find 
\begin{eqnarray}
\partial_{K}K^{l+1+2\eta_\psi}\varphi_{lm}\left(K\right)&=&\lambda_{{\rm p}}z_{l}^{0}\left(2(l+\eta_\psi)+1\right)\nonumber \\
&\times& \int_{K}^{\Lambda}dP\frac{K^{2(l+\eta_\psi)}}{P^{1+l}}\varphi_{lm}\left(P\right),\label{eq:one derivative}
\end{eqnarray}
which leads to the second order differential equation
\begin{eqnarray}
\partial_{K}K^{-2(l+\eta_\psi)}\partial_{K}K^{l+1+2\eta_\psi}\varphi_{lm}\left(K\right)&=&-\lambda_{{\rm p}}\left(2(l+\eta_\psi)+1\right)\nonumber \\
&\times & z_{l}^{0}\frac{\varphi_{lm}\left(P\right)}{K^{1+l}}.\label{eq:dgl}
\end{eqnarray}
From Eq.~\eqref{eq:dgl} also follow the UV and IR boundary conditions
$\left.\partial_{K}K^{l+1+2\eta_\psi}\varphi_{lm}\left(K\right)\right|_{K=\Lambda}=0$
and $\left.K\partial_{K}\varphi_{lm}\left(K\right)\right|_{K=T}=l\varphi_{lm}\left(T\right)$, respectively.
Finally, we use logarithmic variables 
\begin{equation}
\varphi_{lm}\left(K\right)=K^{-1/2-\eta_\psi}f_{lm}\left(\log\frac{K}{\Lambda}\right),
\end{equation}
and the above differential equation takes a particularly simple form of a classical harmonic oscillator problem:
\begin{equation}
\frac{d^{2}f_{lm}\left(x\right)}{dx^{2}}=v_{l}f_{lm}\left(x\right),
\end{equation}
with $v_{l}=\frac{1}{4}\left(2(l+\eta_\psi)+1\right)^{2}-\left(2(l+\eta_\psi)+1\right)\lambda_{{\rm p}}z_{l}^{0}$.
The boundary conditions are $\left.\partial_{x}f_{lm}\left(x\right)\right|_{x=0}=-\frac{2(l+\eta_\psi)+1}{2}f_{lm}\left(0\right)$
in the UV and $\left.\partial_{x}f_{lm}\left(x\right)\right|_{x=x_{T}}=\frac{2(l+\eta_\psi)+1}{2}f_{lm}\left(x_{T}\right)$
with $x_{T}=\log\left(\Lambda/T\right)$ in the IR. If $v_{l}>0$
one cannot simultaneously fulfill both boundary conditions. This changes
once $\nu_{l}$ becomes negative and the solutions of the differential
equation become oscillatory with $f_{lm}\left(x\right)\sim e^{\pm i\sqrt{-v_{l}}x}$.
For $\nu_l<0$, the boundary conditions then determine the transition temperature as
\begin{equation}
    T_{c}=\Lambda\exp\left(-\frac{1}{\sqrt{\left|\nu_l\right|}}\left(\pi-\arctan\frac{4\left(1+2\eta_{\psi}\right)\sqrt{\left|\nu_l\right|}}{\left(1+2\eta_{\psi}\right)^{2}-4\left|\nu_l\right|}\right)\right),
    \label{eq:TcDan}
\end{equation}
which we plot for $l=0$ in Fig.~\ref{fig:TcCurve}. Near the onset of superconductivity this expression simplifies to
\begin{equation}
T_{c}=\Lambda\exp\left(-\frac{D}{\sqrt{\lambda_{{\rm p}}-\lambda_{{\rm p}}^{c}}}\right), \label{eq:thresholdTc}
\end{equation}
where $D=\frac{\pi}{\sqrt{1+2\eta_\psi}}$.
The coupling constant $\lambda_{{\rm p}}$ of Eq.~\eqref{eq:pairing_coupling} must therefore be larger than the  critical value $\lambda_{{\rm p}}^{c}=\frac{1}{4 z_{l}^{0}}\left(2(l+\eta_\psi)+1\right)$, determined from the condition that $\nu_l$ vanishes.
Hence a superconducting
ground state only emerges if $\lambda_{{\rm {\rm p}}}$ of Eq.~\eqref{eq:pairing_coupling}
is larger than $\lambda_{{\rm p}}^{c}$.  The behavior Eq.~\eqref{eq:thresholdTc} for the transition temperature is common to a number of quantum-critical pairing states~\cite{abanov2001,wang2020,Hauck2020}, and generally associated with the spontaneous breaking of conformal symmetry~\cite{Kaplan2009}.

Returning from logarithmic variables to our usual momenta yields
\begin{equation}
\varphi_{lm}\left(P\right)\propto P^{-\frac{1}{2}-\eta_\psi\pm i\delta},\label{eq:powerlaw}
\end{equation}
where the exponent $\delta=\sqrt{|\nu_l}|$ vanishes for $\lambda_{{\rm p}}\rightarrow\lambda_{{\rm p}}^{c}$
from above. It turns out that this is indeed the correct solution
of the full integral Eq.~\eqref{eq:47},  provided the following condition is met:
\begin{eqnarray}
1 & = & 2\lambda_{{\rm p}}\int_{0}^{1}ds\frac{z_{l}\left(s\right)\cos\left(\delta\log s\right)}{\left(s^{2}+1\right)^{\frac{1}{2}+\eta_\psi}s^{-\eta_\psi}}.
\end{eqnarray}
The critical coupling constant for the onset of
pairing with angular momentum $l$ is obtained if one considers $\delta\rightarrow0$
\begin{equation}
  \lambda_{{\rm p},l}^{c}=\frac{1}{2}  \left(\int_{0}^{1}ds\frac{z_{l}\left(s\right)}{\left(s^{2}+1\right)^{\frac{1}{2}+\eta_\psi}s^{-\eta_\psi}}\right)^{-1};
\end{equation}
the ground state is superconducting if $ \lambda_{\rm p}>  \lambda_{{\rm p},l}^{c}$.

\begin{figure}
    \centering
    \includegraphics[width=0.8\linewidth]{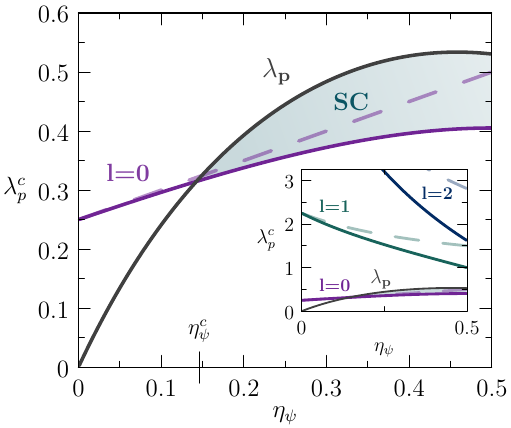}
    \caption{Critical coupling constants $\lambda_{\rm p}^c$ for various angular momenta $l$ for the exact solution (solid lines) and approximation (dashed lines) of the single components superconducting problem that occurs for the pairing interactions $\Upsilon_1$, $\Upsilon_2$, and $\Upsilon_4$. The pairing interaction $\lambda_{\rm p}$ exceeds the critical value for pairing for states with
    angular momentum in space-time $l=0$ and  for $\eta_\psi>0.14628$, yielding  a  superconducting state (SC). The inset shows $\lambda_{\rm p}$ and the critical coupling constants $\lambda_{\rm p}^c$ for higher angular momenta $l$.}
    \label{fig:lambda_crit}
\end{figure}

\begin{figure}
    \centering
\includegraphics[width=0.8\linewidth]{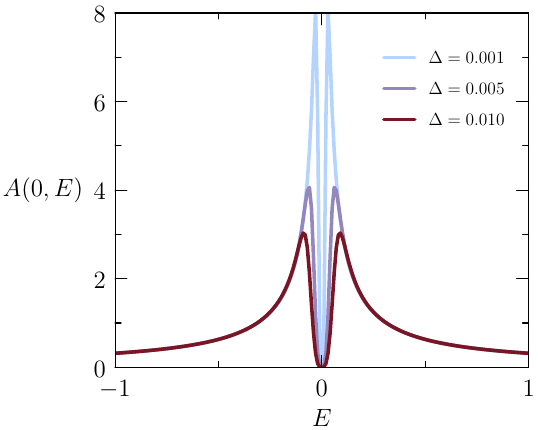}
    \caption{ Single-particle spectral function $A(\boldsymbol{k}, E) = -{\rm Im}\big[ {\rm Tr} G(\boldsymbol{k}, i\omega\to E+i0^+)\big]$  vs energy $E$, evaluated at $\boldsymbol{k}=\boldsymbol{0}$ (Dirac point)
  for the case of $S$ pairing with diagonal self energy Eq.~(\ref{eq:fullselfenergy})
  pairing self energy $\Phi(\boldsymbol{k},\omega) = \Delta 
   (\boldsymbol{k}^2+\omega^2)^{-\frac{1}{4}(1+2\eta_\psi)}$.
  For this plot we chose amplitude $A = 1$, exponent $\eta_\psi = 0.2$,
  pairing amplitude $\Delta = 0.001$ (blue curve), $\Delta = 0.005$ (purple curve)
  and $\Delta = 0.01$ (red curve), showing that the onset of pairing
  rapidly suppresses the spectral function peak while also introducing
  a gap at the Fermi level.
    }
    \label{fig:SCspectral}
\end{figure}

In Fig.~\ref{fig:lambda_crit} we plot $  \lambda_{{\rm p},l}^{c}$ for the various angular
momentum states $l$. Clearly the leading instability is the one 
with $l=0$, while other pairing states require significantly larger
coupling constants. We also compare the critical coupling constant
with the approximate result that follows from the analysis of the
differential equation. For exponents $\eta_\psi$ not too
far from zero is the agreement very good. In the figure we also show
 $\lambda_{{\rm {\rm p}}}$ of Eq.~\eqref{eq:pairing_coupling}. 
We see that only $l=0$ instabilities are allowed and require $\eta_\psi>\eta_\psi^c$, with $\eta_\psi^c$ of Eq.~\eqref{eq:eta_crit}. For all angular momenta $l\geq 1$, the pairing strength is not strong enough to induce higher angular-momentum pairing.  Hence,  if the anomalous dimension of the fermions is sufficiently large,  the ground state of the problem is superconducting and the pairing wave function is isotropic as function of the three momentum ($l=0$). In particular this implies that the pairing state is of even frequency. Thus, for $\Upsilon_1=\gamma^0$ 
 the pairing state is
\begin{equation}
    \Phi\left(P\right)\sim P^{-\frac{1}{2}-\eta_\psi\pm i\delta} 1,
\end{equation}
for $\Upsilon_2=i\gamma^0 \gamma^3$ it holds
\begin{equation}
    \Phi\left(P\right)\sim P^{-\frac{1}{2}-\eta_\psi\pm i\delta} \gamma^3,
\end{equation}
while for 
 $\Upsilon_4=i\gamma^0 \gamma^5$ it holds
\begin{equation}
    \Phi\left(P\right)\sim P^{-\frac{1}{2}-\eta_\psi\pm i\delta} \gamma^5.
\end{equation}
In Fig.~\ref{fig:SCspectral} we show the spectral function on the real frequency axis in the superconducting state and near the onset of pairing that results from this anomalous pairing state. Clearly, states at the Dirac point are gapped by pairing, i.e. the GN-gap due to condensation of the critical boson  is preempted by the onset of superconductivity.

\begin{figure}
    \centering
    \includegraphics[width=0.8\linewidth]{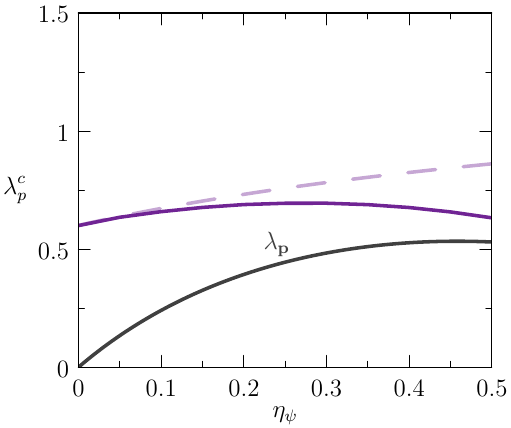}
    \caption{Same as Fig.~\ref{fig:lambda_crit}, but for a pairing interaction $\Upsilon_3$ that leads to the $3\times3$ matrix $Q$ which occurs in the vector component $l_{V}$.   The dashed-purple line corresponds to the threshold coupling obtained from the approximate solution  of the differential equation, while the solid-blue line corresponds to the full solution.  For this interaction the pairing interaction $\lambda_p$ (shown as black line) is always smaller than the threshold coupling for superconductivity $\lambda^c_p$, i.e. no superconductivity emerges.}
    \label{fig:lambda_critV}
\end{figure}

\subsubsection{Vector, Tensor, and Axial Vector pairing states}
In this section we analyze pairing instabilities for the situation where $l_{J}^{rr'}\left(p\right)$ describes coupling in a higher-dimensional state of pairing states and is no-longer  $1\times 1$ matrix. This is important for the pairing interaction $\Upsilon_3=i\gamma^1\gamma^2$, where, according to Table~\ref{tab:pairing_states} the $3\times3$ block $Q$ enters as the only option to yield an attractive interaction. 

 To analyze these multi-component pairing states we need to use the expansion of $l_{J}^{rr'}\left(p\right)$
given in Eq.~\eqref{eq:SHexp}. In analogy to the previous section, we first analyze the solution by approximately transforming the integral equation into a differential
equation. With 
\begin{equation}
 \varphi_{lm}^{r}\left(K\right)=K^{-\tfrac{1}{2}-\eta_\psi}f_{lm}^{r}\left(\log K\right),   
\end{equation}
 it follows that  
\begin{equation}
 \frac{d^{2}f_{lm}^{r}\left(x\right)}{dx^{2}}=\sum_{r'l'm'}V_{lm,l'm'}^{rr'}f_{l'm'}^{r'}\left(x\right),   
\end{equation}
 where we defined
\begin{equation}
     V_{lm,l'm'}^{rr'}=\frac{\left(2(l+\eta_\psi)+1\right)^{2}}{4} \left(\delta_{rr'}\delta_{ll'}\delta_{mm'}-\lambda_{{\rm p}} U_{lm,l'm'}^{(0)rr'}\right),
\end{equation}
with
\begin{equation}
   U_{lm,l'm'}^{(0)rr'} =\frac{4\tau_{i}z_{l}^{0}}{\left(2(l+\eta_\psi)+1\right)}M_{lm,l'm'}^{rr'}.
\end{equation}
Pairing corresponds to the smallest eigenvalue of $V$ crossing zero, i.e. the largest eigenvalue of $U^{(0)}$ reaching $1/\lambda_{{\rm p}}$. The result for the the $3\times 3$ block $l_V=Q$ of Table~\ref{tab:pairing_states} is shown in Fig.~\ref{fig:lambda_critV}. We find that the pairing strength $\lambda_{\rm p}$ never crosses the critical value. 

To check this result, we solve the linearized gap equation more carefully and find that
 the ansatz
\begin{equation}
\varphi_{lm}^{r}\left(P\right)=A_{lm}^{r}P^{-\frac{1}{2}-\eta_\psi\pm i\beta},
\end{equation}
 solves the integral equation, provided the largest eigenvalue of
the matrix 
\begin{equation}
U_{lm,l'm'}^{rr'}=2\tau_{i}M_{lm,l'm'}^{rr'}\int_{0}^{1}ds\frac{z_{l}\left(s\right)}{\left(s^{2}+1\right)^{\frac{1}{2}+\eta_\psi}s^{-\eta_\psi}}.\label{eq:U_full}
\end{equation}
equals $\lambda_{{\rm p}}^{-1}$. In Fig.~\ref{fig:lambda_critV} we show for the $3\times3$ matrix $Q$ that enters
in the vector component $l_{V}$ the approximate
solution from the differential equation as well as the full solution
using $U$ of Eq.~\eqref{eq:U_full} as function of the anomalous exponent
$\eta_{\psi}$. We find that the condition $\lambda_{{\rm p}}>\lambda_{{\rm p}}^{c}$
can not be fulfilled, i.e. critical fluctuations of a boson coupled via $\Upsilon_3=i\gamma^{1}\gamma^{2}$
will not give rise to a stable superconducting state. The complexity of the multi-component pairing state leads to an enhanced threshold coupling, which is always larger than $\lambda_p$, including  for $l=0$.

\subsection{Algebraic conditions for pairing}
Our results for pairing can be expressed very efficiently as algebraic conditions on the matrices $\Gamma^r_J$ that, according to Eq.~\eqref{eq:Delta_expansion}, govern the spinor structure of the pairing state. To this end, we generalize to $n_\gamma$-component spinors and to bosons with $m_\phi$ degenerate components, described by $\Upsilon_i$ ($i=1,\dots,m_\phi$), all sharing the same parity under time reversal $\tau$. In this framework, the matrix $L^{rr'}_{JJ'}(p)$ of Eq.~\eqref{eq:matrix_L} takes the form
\begin{equation}
L^{rr'}_{JJ'}\left(p\right)=\frac{p_{\mu}p_{\nu}}{m_{\phi}n_{\gamma}p^{2}}\sum_{i=1}^{m_{\phi}}{\rm tr}\left(\Upsilon_{i}\gamma^{0}\gamma^{\mu}\Gamma^r_{J}\gamma^{\nu}\gamma^{0}\Upsilon_{i}\Gamma^r_{J'}\right).
\label{eq:L_most_general}
\end{equation}
It is now an $n_\gamma^2\times n_\gamma^2$ matrix. 
The analysis of the linearized pairing problem implies that superconductivity becomes possible if  $L\left(p\right)$
has one-dimensional irreducible sub-blocks with $\left(L\right)_{{\rm sub\,block}}=\tau$. One can easily show that this condition is fulfilled for the $(r,J)$ that obey
\begin{eqnarray}
\left[\Gamma^r_{J},\alpha_{i}\right] & = & 0\,\,\,{\rm for}\,\,i=1,2,
\label{eq:Cond1}
\end{eqnarray}
and 
\begin{equation}
\Gamma^r_{J}\Upsilon_{i}=\tau \Upsilon_{i}\Gamma^r_{J}\,\,\,{\rm for}\,\,i=1\cdots m_{\phi}.
\label{eq:Cond2}
\end{equation}
Hence, for $\tau=-1$ it holds that the pairing matrix and the $\Upsilon_{i}$
anti-commute $\left\{ \Gamma^r_{J},\Upsilon_{i}\right\} =0$ while they
commute for $\tau=+1$, $\left[\Gamma^r_{J},\Upsilon_{i}\right]=0$.
An allowed pairing state
must also obey Pauli principle, which for  an angular-momentum $l=0$ state implies 
\begin{equation}
(\Gamma_{J}u_{T})^T=-\Gamma_{J}u_{T}.
\label{eq:Cond3}
\end{equation}
The three algebraic conditions Eqn.~\eqref{eq:Cond1}, \eqref{eq:Cond2}, and \eqref{eq:Cond3} for  $\Gamma^r_J$ allow for an easy determination of superconductivity in a generic Dirac problem coupled to arbitrary critical bosons.
Interestingly, the second condition, Eq.~\eqref{eq:Cond2}, was recently obtained in the study of the leading pairing instability of twisted bilayer graphene in the extreme flat-band limit~\cite{Christos2023nodal}, i.e. not in the Dirac regime, suggesting that it may in fact be of more general relevance. This point will be elaborated further in Ref.~\cite{Stangier2025}. By contrast, the first condition, Eq.~\eqref{eq:Cond1}, is specific to the Dirac theory.

\subsection{Absence of pairing for two-component Dirac spinors}
It is instructive to apply our formalism and analyze for the possibility
of pairing near the GN-critical point for a two-component Dirac spinor,
i.e. the simplest Dirac theory in $2+1$ dimensions. To this end we make, without restrictions, the choice
\begin{equation}
\alpha_{1}=\sigma_{1},\,\,\alpha_{2}=\sigma_{2},
\end{equation}
coupled to the only mass-generating term with 
\begin{equation}
\Upsilon=\sigma_{3}.
\end{equation}
The $\sigma_{i}$ are the usual Pauli matrices. If the fermions are
spin-less the unitary component of time reversal is $u_{T}=\sigma_{0}$
(the unit matrix) and $\Upsilon$ is time-reversal even, i.e. $\tau=1$.
If $\sigma_{i}$ stand for spin, then $u_{T}=i\sigma^{y}$ and $\Upsilon$
is time-reversal odd with $\tau=-1$. The complete set of Hermitian
$2\times2$ matrices is given by $\Gamma_{J}=\sigma_{J}$ with $J=0\cdots3$.
This immediately allows to determine the matrix 
\begin{equation}
L=\left(\begin{array}{cc}
l_{S} & 0\\
0 & l_{V}
\end{array}\right),
\end{equation}
where $l_{S}=1$ and $l_{V}=-Q$ with the $3\times3$ matrix used
earlier. As before, $l_{V}$ will not induce pairing. The pairing
wave function in the $l_{S}$ channel is $\Phi\left(k\right)=\chi\left(k\right)\sigma_{0}$.
It seems that a time-reversal even critical boson might cause superconductivity
since $\tau l_{S}=+1$. However, due to $u_{T}=\sigma_{0}$ follows
$\Delta\left(k\right)=\chi\left(k\right)\sigma_{0}$ and the function
$\chi\left(k\right)$ must be odd in $k$ to comply with Pauli principle.
Even for the smallest angular momentum $l=1$ the threshold value
for the pairing interaction is too large. Pauli principle would be
consistent with $l=0$ for spin-full fermions where $\Delta\left(k\right)=\chi\left(k\right)i\sigma_{2}$
corresponds to a singlet state. However, in this case $\tau l_{S}=-1$
and the interaction is repulsive. Hence, no superconductivity emerges in a two-component Dirac spinor.  

We can easily come to the same conclusion using the three algebraic conditions Eqn.~\eqref{eq:Cond1}, \eqref{eq:Cond2}, and \eqref{eq:Cond3}: Given our choice for $\alpha_1$ and $\alpha_2$, the only $2\times 2$ matrix that commutes with both, i.e. that is a candidate for $\Gamma_J$, is $\sigma_0$. $\sigma_0$ does not anti-commute with $\Upsilon$, i.e. there is no attractive coupling for time-reversal odd bosons. It commutes with $\Upsilon$ and could serve as pairing state for a time-reversal even boson. However, in this case it holds that $u_T=\sigma_0$ and $\sigma_0 u_T$ is not antisymmetric. Hence, we arrive at the same conclusion that no pairing state is allowed. More complex
spinor structures  are necessary for pairing near
the GN critical point to emerge. 

\subsection{Superconducting order-parameter fluctuations}
The solution of the pairing problem yields an
instability temperature $T_{c}$. Within the large-$N$ approach used
here, superconducting order-parameter fluctuation are suppressed and
and $T_{c}$ corresponds to the actual phase transition temperature.
For any finite $N$ order-parameter fluctuations are of course important
and ultimately give rise to a Berezinskii--Kosterlitz--Thouless
(BKT) transition~\cite{Berezinskii1971,Kosterlitz1973,Halperin1979}. In what follows we show that such fluctuations will
give rise to corrections to $T_{c}$ of order unity, but that ultimately
the BKT transition temperature remains finite and is of the order of magnitude of $T_{c}$.

To analyze BKT physics we need to determine the phase stiffness of
the problem. The stiffness then determines the behavior of the two-dimensional
classical problem at finite temperature with XY-action 
\begin{equation}
S_{{\rm XY}}=\frac{\rho_{s}^{\left(0\right)}}{2}\int d^{2}x\left(\nabla\theta-2e\boldsymbol{A}\right)^{2},
\end{equation}
where $\theta$ is the phase of the superconducting order parameter and $\boldsymbol{A}$ the electromagnetic vector potential.  Order-parameter fluctuations renormalize $\rho_{s}^{\left(0\right)}\rightarrow\rho_{s}$
and the BKT transition temperature follows from the celebrated condition~\cite{Kosterlitz1973,Halperin1979}
\begin{equation}
\rho_{s}\left(T_{{\rm BKT}}\right)=\frac{2}{\pi}T_{{\rm BKT}}.
\end{equation}
The determination of the superfluid stiffness is a nontrivial analysis, which requires the determination of current vertex corrections; a consequence of the strong momentum dependence of the single-particle self energy. This is rather different from the usual analysis in Eliashberg-type theories with momentum-independent self energies; for a detailed discussion see Ref.~\cite{Raines2024}. Qualitative  understanding
can, however, be obtained using the Ferrell-Glover-Tinkham (FGT) sum
rule of the real part $\sigma'(\omega)$ of the optical conductivity~\cite{Ferrell1958,Tinkham1959}:
\begin{equation}
\frac{\omega_{p}^{2}}{4}=\int_{-\infty}^{\infty}d\omega\sigma'\left(\omega\right),
\end{equation}
with plasma frequency $\omega_{p}$. For a Dirac particle with upper
cut off $\Lambda$ the sum rule becomes~\cite{Sabio2008}: 
\begin{equation}
\int_{-\Lambda}^{\Lambda}d\omega\sigma'\left(\omega\right)=\frac{N}{4}\Lambda.
\end{equation}
In the superconducting state it
holds for the real part of the optical conductivity
\begin{equation}
\sigma'_{{\rm sc}}\left(\omega\right)=\pi e^{2}\rho_{s}\delta\left(\omega\right)+\tilde{\sigma}_{{\rm ns}}\left(\omega\right),
\end{equation}
where $\tilde{\sigma}_{{\rm ns}}\left(\omega\right)$ essentially
equals the normal state conductivity for $\left|\omega\right|>T$
and is due to inter-band transitions. We expect it to vanish at low
frequencies, below the pairing gap $\Delta$. We further expect that this
gap is parametrically determined by the mean-field transition temperature:
$\Delta\sim T_{c}$, where the coefficient is of the order
of unity but typically somewhat larger. In the normal state,  hydrodynamic arguments~\cite{Fritz2008,Inkof2020} fix the conductivity for $\omega<T$ to
a Drude contribution:
\begin{equation}
\sigma'_{{\rm ns}}\left(\omega\right)=\frac{\sigma_{0}}{1+\left(\omega\tau\right)^{2}}+\tilde{\sigma}_{{\rm ns}}\left(\omega\right).
\end{equation}
For the scattering rate we expect at a critical point $\tau^{-1}=\alpha^{*}T$,
with fixed point value of the interaction $\alpha^{*}$. Then follows from the FGT sum rule 
for $\Delta\tau\approx1$ that 
\begin{equation}
\pi\frac{e^{2}}{\hhbar}\rho_{s}\approx\sigma_{0}\Delta.
\end{equation}
In the usual weakly disordered Fermi liquid it holds $\sigma_{0}=\frac{\omega_{{\rm p}}^{2}}{4\pi}\tau$
and we obtain for the stiffness in the superconducting state $\rho_{s}=\frac{n}{m}\frac{\pi}{4}\tau\Delta$ with $\omega_{p}^{2}=\pi\frac{e^{2}}{\hhbar}\frac{n}{m}$.  For $\frac{n}{m}\frac{\pi}{4}\tau\approx E_{F}\tau\gg1$ the stiffness, albeit reduced from its clean limit, is large
compared to $\Delta$ and phase fluctuations are relevant only
in a very narrow regime near $T_{c}$. Then  $T_{{\rm BKT}}$ is smaller
than, but very close to $T_{c}$~\cite{Halperin1979}. In our problem we expect at the
GN-fixed point  for the conductivity
\begin{equation}
\sigma_{0}=N\zeta_{*}\frac{e^{2}}{h},\label{eq:cond_dc}
\end{equation}
where $\zeta_{*}$ should be a dimensionless number of order unity.
This implies 
\begin{equation}
\rho_{s}=N\zeta_{*}\frac{\Delta}{2\pi^{2}}.
\end{equation}
At large $N$ the stiffness is obviously large and there are no phase
fluctuations. Considering $N=1$ follows that $\rho_{\ast}\sim\Delta$
and  phase fluctuations are strong near the transition temperature. Nevertheless. 
$\left(T_{{\rm c}}-T_{{\rm BKT}}\right)/T_{{\rm BKT}}$ of order unity since the stiffness scale is set by $T_c$.
Hence, ultimately $T_{{\rm BKT}}$ should be of the same order as
$T_{c}$. The stiffness at $T=0$ can be understood in a similar way. Although the Drude peak is now absent, a nearly frequency-independent contribution to the optical conductivity from inter-band transitions persists down to the lowest frequencies and is again expected to be of order $Ne^2/h$. In other words, the stiffness is set by the zero-temperature gap. We therefore expect superconductivity to remain stable at sufficiently low temperatures, even for finite $N$. 

\begin{figure}
    \centering
    \includegraphics[width=0.9\linewidth]{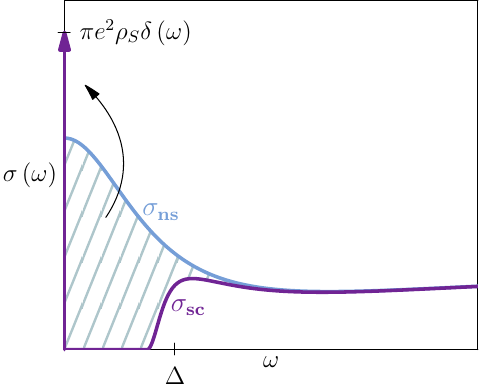}
    \caption{The transfer of spectral weight in the optical conductivity from the normal state behavior (light purple) to the superconducting state (dark purple) determine the superfluid stiffness $\rho_S$ and is determined by the gap energy scale $\Delta$ and the normal state D.C. conductivity $\sigma_0$. For the latter we expect $\sigma_0\sim N e^2/\hhbar$ which implies $\rho_S\sim N \Delta$, i.e. phase fluctuations are important at finite $N$ and change the transition temperature compared to its value obtained from the large-$N$ theory. }
    \label{fig:Sigma}
\end{figure}

\section{Summary}
In summary, we analyzed the possibility of superconductivity due to
critical fluctuations at the Gross-Neveu critical point for two-dimensional
massless Dirac fermions at neutrality, coupled to a bosonic mode by
Yukawa coupling. For the bosonic mode we analyzed the four options
that induce, upon bose condensation an isotropic gap in the fermion
spectrum, see Eq.~\eqref{eq:coupling_matrices}. In its normal state, the theory is identical
to the one developed by Kim et al.~\cite{Kim2021} and yields anomalous dimensions
$\eta_{\psi}$ and $\eta_{\phi}$ for the Dirac fermions and critical
bosons of the problem. Importantly, one then finds values for $\eta_{\psi}$ that are not parametrically small, in distinction to the usual expansions in $1/N$ (with fermion
flavor $N$) or in $\epsilon=3-d$. We then generalized the approach of Ref.~\cite{Kim2021} to the superconducting state and analyzed the linearized
gap equation of the pairing problem. We find that superconductivity emerges once the anomalous fermion dimension exceeds a critical threshold. Strikingly, our analysis shows that superconductivity is absent in the regime of well-defined quasiparticles, but appears when the quasiparticles become ill-defined. 
We considered four distinct pairing interactions and found that superconductivity
emerges in three out of the four cases. We further list easy-to-analyze algebraic conditions (Eqn.~\eqref{eq:Cond1}, \eqref{eq:Cond2}, and \eqref{eq:Cond3}) that allow one to determine  superconducting states for a generic Gross-Neveu theory with arbitrary spinor components and critical boson modes.

Since our theory was formulated
for a generic representation of the Dirac matrices, it is instructive
to discuss it in the context of a specific realization. To this end we consider
\begin{equation}
\alpha_{1}=-\tau_{2}\sigma_{2},\,\,\alpha_{2}=\tau_{2}\sigma_{1},\,\,\alpha_{3}=-\tau_{1}\sigma_{0}\,{\rm and}\,\,\beta=\tau_{3}\sigma_{0},\label{eq:alpha_matrices}
\end{equation}
discussed in Refs.~\cite{Liu2010,Palle2024b}. Here $\tau_{a}$ stands for two
orbitals of opposite parity while $\sigma_{b}$ acts in spin space.
For this problem it holds that the parity operation is $u_{P}=\beta$
while time reversal is given as $u_{T}=i\tau_{3}\sigma_{2}$.  The pairing states for the four distinct Yukawa couplings, i.e.
the interaction terms $\sim g\psi^{\dagger}\Upsilon_i\psi\phi$ in the
action can then easily be analyzed with the help of Table~\ref{tab:pairing_states}. The coupling
\begin{equation}
\Upsilon_1=\gamma^{0}=\tau_{3}\sigma_{0},
\end{equation}
which is even under inversion and time reversal and corresponds to
an excitation in the charge channel, describes 
an orbital fluctuation. For the superconducting gap function induced
by this coupling follows
\begin{equation}
\Delta\left(k\right)=\chi\left(K\right)u_{T}=-i\chi\left(K\right)\tau_{3}\sigma_{2},
\end{equation}
i.e. we obtain a spin-singlet, orbital triplet that is out of phase
for the two orbitals. The coupling 
\begin{equation}
\Upsilon_2=i\gamma^{0}\gamma^{3}=-\tau_{1}\sigma_{0},
\end{equation}
which is odd under parity and time-reversal, and hence corresponds
to a toroidal moment. It is trivial in spin space but describes transitions
between the orbital states that amount to orbital currents, i.e. some
form of loop currents of the two-orbital problem. For the superconducting
gap function induced by this coupling follows
\begin{equation}
\Delta\left(k\right)=\chi\left(K\right)\gamma^{3}u_{T}=-\chi\left(K\right)\tau_{1}\sigma_{2}.
\end{equation}
Hence, we obtain a different spin-singlet / orbital triplet state.
The coupling 
\begin{equation}
\Upsilon_4=i\gamma^{0}\gamma^{5}=\tau_{2}\sigma_{2},
\end{equation}
is also odd under parity and time-reversal and describes spin-orbit
entangled toroidal moment. For the superconducting gap function induced
by this coupling follows
\begin{equation}
\Delta\left(k\right)=\chi\left(K\right)\gamma^{5}u_{T}=i\chi\left(K\right)\tau_{2}\sigma_{1}.
\end{equation}
This corresponds to an orbital singlet and spin triplet state. Finally
the absence of superconductivity due to 
\begin{equation}
\Upsilon_3=i\gamma^{1}\gamma^{2}=-\tau_{0}\sigma_{3},
\end{equation}
which is odd under time reversal and even under parity and describes
with Eq.~\eqref{eq:alpha_matrices} spin-ferromagnetic fluctuations,
shows that not all Gross-Neveu interactions serve equally efficient
as pairing glue. This interaction is most attractive in the channel
with the three-component gap function
\begin{eqnarray}
\Delta\left(k\right) & = & \sum_{\mu=0}^{2}\chi^{\mu}\left(K\right)\gamma^{\mu}u_{T} , \\
 & = & -i\chi^{0}\left(K\right)\tau_{0}\sigma_{2}-\chi^{1}\left(K\right)\tau_{2}\sigma_{0}+i\chi^{0}\left(K\right)\tau_{2}\sigma_{3},\nonumber
\end{eqnarray}
which transforms under a three-dimensional
irreducible representation. It is a combination of an orbital triplet
and spin singlet, with amplitude $\chi^{0}\left(K\right)$ with two
orbital singlets / spin triplets, with amplitudes $\chi^{1,2}$. While
superconducting fluctuations may be sizable, the pairing
interaction does not reach the threshold value in this pairing channel.

The choice of Dirac matrices in Eq.~\eqref{eq:alpha_matrices} illustrates
the rich physics of unconventional pairing states that emerge upon
exchanging critical mass-generating bosons in Dirac systems. Our formalism can be readily applied to generic representations of Dirac matrices, allowing one to analyze whether, for a given system, the relevant collective boson can induce pairing and to determine the resulting pairing symmetry. In a subsequent publication~\cite{Stangier2025} 
the approach  will be applied to the case of $16\times16$ Dirac matrices that one encounters in the context of twisted bilayer graphene~\cite{Bultinck2020,Liu2021,Parthenios2023,Biedermann2025,Hawashin2025,Huang2025} and to the $8\times 8$ Dirac spinors that describe AB-BA
stacked twisted double-layer WSe$_{2}$ at filling $\nu=2$\cite{Angeli2021,Pan2023,Foutty2023,Ma2024,Tolosa2025}.

Let us also comment on the emergence of superconductivity, induced
by time-reversal and parity-odd fluctuations that do not break translation.
Recently it was pointed out that such a collective boson is unable
to induce superconductivity in the critical regime~\cite{Shi2023loop,Palle2024,Palle2024b,Schultz2025}. We emphasise that these restrictions apply only to systems without spin-orbit coupling and  therefore do not pertain  to our problem.

Finally, we discussed the role of superconducting order-parameter fluctuations beyond the large-$N$ limit, concluding that such fluctuations are likely to reduce the transition temperature but should not destroy the superconducting state identified here. Solving the pairing problem below $T_c$,  addressing the robustness of pairing against disorder and externally applied magnetic fields, and elucidating the connection to topological superconductivity due to charged skyrmions, as discussed in Refs.~\cite{Wiegmann1999,Abanov2000theta,Grover2008,Christos2020,Khalaf2021charged,Ledwith2021strong}, are among the important open problems that emerge from our findings.

\begin{acknowledgments}
We are grateful to Ehud Altman, Andrey V. Chubukov, Laura Classen, Ilya Esterlis, Jaewon Kim,  Avraham Klein, Elio K\"onig, Grgur Palle, Nikolaos Parthenios, Jonathan Ruhman, and Mathias Scheurer for helpful discussions. This work was supported
by the German Research Foundation TRR 288-422213477 ELASTO-Q-MAT,
 B01 (V.C.S. and J.S.) and grant  SFI-MPS-
NFS-00006741-05 from the Simons Foundation (J.S.).  
D.E.S. acknowledges support from the National Science Foundation under Grant PHY-2208036.   
\end{acknowledgments}

\bibliography{refs}

\begin{thebibliography}{99}%
\makeatletter
\providecommand \@ifxundefined [1]{%
 \@ifx{#1\undefined}
}%
\providecommand \@ifnum [1]{%
 \ifnum #1\expandafter \@firstoftwo
 \else \expandafter \@secondoftwo
 \fi
}%
\providecommand \@ifx [1]{%
 \ifx #1\expandafter \@firstoftwo
 \else \expandafter \@secondoftwo
 \fi
}%
\providecommand \natexlab [1]{#1}%
\providecommand \enquote  [1]{``#1''}%
\providecommand \bibnamefont  [1]{#1}%
\providecommand \bibfnamefont [1]{#1}%
\providecommand \citenamefont [1]{#1}%
\providecommand \href@noop [0]{\@secondoftwo}%
\providecommand \href [0]{\begingroup \@sanitize@url \@href}%
\providecommand \@href[1]{\@@startlink{#1}\@@href}%
\providecommand \@@href[1]{\endgroup#1\@@endlink}%
\providecommand \@sanitize@url [0]{\catcode `\\12\catcode `\$12\catcode `\&12\catcode `\#12\catcode `\^12\catcode `\_12\catcode `\%12\relax}%
\providecommand \@@startlink[1]{}%
\providecommand \@@endlink[0]{}%
\providecommand \url  [0]{\begingroup\@sanitize@url \@url }%
\providecommand \@url [1]{\endgroup\@href {#1}{\urlprefix }}%
\providecommand \urlprefix  [0]{URL }%
\providecommand \Eprint [0]{\href }%
\providecommand \doibase [0]{http://dx.doi.org/}%
\providecommand \selectlanguage [0]{\@gobble}%
\providecommand \bibinfo  [0]{\@secondoftwo}%
\providecommand \bibfield  [0]{\@secondoftwo}%
\providecommand \translation [1]{[#1]}%
\providecommand \BibitemOpen [0]{}%
\providecommand \bibitemStop [0]{}%
\providecommand \bibitemNoStop [0]{.\EOS\space}%
\providecommand \EOS [0]{\spacefactor3000\relax}%
\providecommand \BibitemShut  [1]{\csname bibitem#1\endcsname}%
\let\auto@bib@innerbib\@empty
\bibitem [{\citenamefont {Kim}\ \emph {et~al.}(2021)\citenamefont {Kim}, \citenamefont {Altman},\ and\ \citenamefont {Cao}}]{Kim2021}%
  \BibitemOpen
  \bibfield  {author} {\bibinfo {author} {\bibfnamefont {J.}~\bibnamefont {Kim}}, \bibinfo {author} {\bibfnamefont {E.}~\bibnamefont {Altman}}, \ and\ \bibinfo {author} {\bibfnamefont {X.}~\bibnamefont {Cao}},\ }\href {https://doi.org/10.1103/PhysRevB.103.L081113} {\bibfield  {journal} {\bibinfo  {journal} {Physical Review B}\ }\textbf {\bibinfo {volume} {103}},\ \bibinfo {pages} {L081113} (\bibinfo {year} {2021})}\BibitemShut {NoStop}%
\bibitem [{\citenamefont {Wehling}\ \emph {et~al.}(2014)\citenamefont {Wehling}, \citenamefont {Black-Schaffer},\ and\ \citenamefont {Balatsky}}]{Wehling2014}%
  \BibitemOpen
  \bibfield  {author} {\bibinfo {author} {\bibfnamefont {T.~O.}\ \bibnamefont {Wehling}}, \bibinfo {author} {\bibfnamefont {A.~M.}\ \bibnamefont {Black-Schaffer}}, \ and\ \bibinfo {author} {\bibfnamefont {A.~V.}\ \bibnamefont {Balatsky}},\ }\href {https://doi.org/10.1080/00018732.2014.927109} {\bibfield  {journal} {\bibinfo  {journal} {Advances in Physics}\ }\textbf {\bibinfo {volume} {63}},\ \bibinfo {pages} {1} (\bibinfo {year} {2014})}\BibitemShut {NoStop}%
\bibitem [{\citenamefont {Vafek}\ and\ \citenamefont {Vishwanath}(2014)}]{Vafek2014}%
  \BibitemOpen
  \bibfield  {author} {\bibinfo {author} {\bibfnamefont {O.}~\bibnamefont {Vafek}}\ and\ \bibinfo {author} {\bibfnamefont {A.}~\bibnamefont {Vishwanath}},\ }\href {https://doi.org/10.1146/annurev-conmatphys-031113-133841} {\bibfield  {journal} {\bibinfo  {journal} {Annu. Rev. Condens. Matter Phys.}\ }\textbf {\bibinfo {volume} {5}},\ \bibinfo {pages} {83} (\bibinfo {year} {2014})}\BibitemShut {NoStop}%
\bibitem [{\citenamefont {Boyack}\ \emph {et~al.}(2021)\citenamefont {Boyack}, \citenamefont {Yerzhakov},\ and\ \citenamefont {Maciejko}}]{Boyack2021}%
  \BibitemOpen
  \bibfield  {author} {\bibinfo {author} {\bibfnamefont {R.}~\bibnamefont {Boyack}}, \bibinfo {author} {\bibfnamefont {H.}~\bibnamefont {Yerzhakov}}, \ and\ \bibinfo {author} {\bibfnamefont {J.}~\bibnamefont {Maciejko}},\ }\href {https://doi.org/10.1140/epjs/s11734-021-00069-1} {\bibfield  {journal} {\bibinfo  {journal} {The European Physical Journal Special Topics}\ }\textbf {\bibinfo {volume} {230}},\ \bibinfo {pages} {979} (\bibinfo {year} {2021})}\BibitemShut {NoStop}%
\bibitem [{\citenamefont {Novoselov}\ \emph {et~al.}(2004)\citenamefont {Novoselov}, \citenamefont {Geim}, \citenamefont {Morozov}, \citenamefont {Jiang}, \citenamefont {Zhang}, \citenamefont {Dubonos}, \citenamefont {Grigorieva},\ and\ \citenamefont {Firsov}}]{novoselov2004}%
  \BibitemOpen
  \bibfield  {author} {\bibinfo {author} {\bibfnamefont {K.~S.}\ \bibnamefont {Novoselov}}, \bibinfo {author} {\bibfnamefont {A.~K.}\ \bibnamefont {Geim}}, \bibinfo {author} {\bibfnamefont {S.~V.}\ \bibnamefont {Morozov}}, \bibinfo {author} {\bibfnamefont {D.-e.}\ \bibnamefont {Jiang}}, \bibinfo {author} {\bibfnamefont {Y.}~\bibnamefont {Zhang}}, \bibinfo {author} {\bibfnamefont {S.~V.}\ \bibnamefont {Dubonos}}, \bibinfo {author} {\bibfnamefont {I.~V.}\ \bibnamefont {Grigorieva}}, \ and\ \bibinfo {author} {\bibfnamefont {A.~A.}\ \bibnamefont {Firsov}},\ }\href {https://www.science.org/doi/10.1126/science.1102896} {\bibfield  {journal} {\bibinfo  {journal} {Science}\ }\textbf {\bibinfo {volume} {306}},\ \bibinfo {pages} {666} (\bibinfo {year} {2004})}\BibitemShut {NoStop}%
\bibitem [{\citenamefont {Castro~Neto}\ \emph {et~al.}(2009)\citenamefont {Castro~Neto}, \citenamefont {Guinea}, \citenamefont {Peres}, \citenamefont {Novoselov},\ and\ \citenamefont {Geim}}]{castro2009}%
  \BibitemOpen
  \bibfield  {author} {\bibinfo {author} {\bibfnamefont {A.~H.}\ \bibnamefont {Castro~Neto}}, \bibinfo {author} {\bibfnamefont {F.}~\bibnamefont {Guinea}}, \bibinfo {author} {\bibfnamefont {N.~M.~R.}\ \bibnamefont {Peres}}, \bibinfo {author} {\bibfnamefont {K.~S.}\ \bibnamefont {Novoselov}}, \ and\ \bibinfo {author} {\bibfnamefont {A.~K.}\ \bibnamefont {Geim}},\ }\href {\doibase 10.1103/RevModPhys.81.109} {\bibfield  {journal} {\bibinfo  {journal} {Rev. Mod. Phys.}\ }\textbf {\bibinfo {volume} {81}},\ \bibinfo {pages} {109} (\bibinfo {year} {2009})}\BibitemShut {NoStop}%
\bibitem [{\citenamefont {Li}\ \emph {et~al.}(2010)\citenamefont {Li}, \citenamefont {Luican}, \citenamefont {Lopes~dos Santos}, \citenamefont {Castro~Neto}, \citenamefont {Reina}, \citenamefont {Kong},\ and\ \citenamefont {Andrei}}]{li2010}%
  \BibitemOpen
  \bibfield  {author} {\bibinfo {author} {\bibfnamefont {G.}~\bibnamefont {Li}}, \bibinfo {author} {\bibfnamefont {A.}~\bibnamefont {Luican}}, \bibinfo {author} {\bibfnamefont {J.}~\bibnamefont {Lopes~dos Santos}}, \bibinfo {author} {\bibfnamefont {A.}~\bibnamefont {Castro~Neto}}, \bibinfo {author} {\bibfnamefont {A.}~\bibnamefont {Reina}}, \bibinfo {author} {\bibfnamefont {J.}~\bibnamefont {Kong}}, \ and\ \bibinfo {author} {\bibfnamefont {E.}~\bibnamefont {Andrei}},\ }\href {https://doi.org/10.1038/nphys1463} {\bibfield  {journal} {\bibinfo  {journal} {Nature physics}\ }\textbf {\bibinfo {volume} {6}},\ \bibinfo {pages} {109} (\bibinfo {year} {2010})}\BibitemShut {NoStop}%
\bibitem [{\citenamefont {Cao}\ \emph {et~al.}(2018{\natexlab{a}})\citenamefont {Cao}, \citenamefont {Fatemi}, \citenamefont {Demir}, \citenamefont {Fang}, \citenamefont {Tomarken}, \citenamefont {Luo}, \citenamefont {Sanchez-Yamagishi}, \citenamefont {Watanabe}, \citenamefont {Taniguchi}, \citenamefont {Kaxiras} \emph {et~al.}}]{cao2018}%
  \BibitemOpen
  \bibfield  {author} {\bibinfo {author} {\bibfnamefont {Y.}~\bibnamefont {Cao}}, \bibinfo {author} {\bibfnamefont {V.}~\bibnamefont {Fatemi}}, \bibinfo {author} {\bibfnamefont {A.}~\bibnamefont {Demir}}, \bibinfo {author} {\bibfnamefont {S.}~\bibnamefont {Fang}}, \bibinfo {author} {\bibfnamefont {S.~L.}\ \bibnamefont {Tomarken}}, \bibinfo {author} {\bibfnamefont {J.~Y.}\ \bibnamefont {Luo}}, \bibinfo {author} {\bibfnamefont {J.~D.}\ \bibnamefont {Sanchez-Yamagishi}}, \bibinfo {author} {\bibfnamefont {K.}~\bibnamefont {Watanabe}}, \bibinfo {author} {\bibfnamefont {T.}~\bibnamefont {Taniguchi}}, \bibinfo {author} {\bibfnamefont {E.}~\bibnamefont {Kaxiras}},  \emph {et~al.},\ }\href {https://doi.org/10.1038/nature26154} {\bibfield  {journal} {\bibinfo  {journal} {Nature}\ }\textbf {\bibinfo {volume} {556}},\ \bibinfo {pages} {80} (\bibinfo {year} {2018}{\natexlab{a}})}\BibitemShut {NoStop}%
\bibitem [{\citenamefont {Cao}\ \emph {et~al.}(2018{\natexlab{b}})\citenamefont {Cao}, \citenamefont {Fatemi}, \citenamefont {Fang}, \citenamefont {Watanabe}, \citenamefont {Taniguchi}, \citenamefont {Kaxiras},\ and\ \citenamefont {Jarillo-Herrero}}]{cao2018b}%
  \BibitemOpen
  \bibfield  {author} {\bibinfo {author} {\bibfnamefont {Y.}~\bibnamefont {Cao}}, \bibinfo {author} {\bibfnamefont {V.}~\bibnamefont {Fatemi}}, \bibinfo {author} {\bibfnamefont {S.}~\bibnamefont {Fang}}, \bibinfo {author} {\bibfnamefont {K.}~\bibnamefont {Watanabe}}, \bibinfo {author} {\bibfnamefont {T.}~\bibnamefont {Taniguchi}}, \bibinfo {author} {\bibfnamefont {E.}~\bibnamefont {Kaxiras}}, \ and\ \bibinfo {author} {\bibfnamefont {P.}~\bibnamefont {Jarillo-Herrero}},\ }\href {https://doi.org/10.1038/nature26160} {\bibfield  {journal} {\bibinfo  {journal} {Nature}\ }\textbf {\bibinfo {volume} {556}},\ \bibinfo {pages} {43} (\bibinfo {year} {2018}{\natexlab{b}})}\BibitemShut {NoStop}%
\bibitem [{\citenamefont {Andrei}\ and\ \citenamefont {MacDonald}(2020)}]{andrei2020}%
  \BibitemOpen
  \bibfield  {author} {\bibinfo {author} {\bibfnamefont {E.~Y.}\ \bibnamefont {Andrei}}\ and\ \bibinfo {author} {\bibfnamefont {A.~H.}\ \bibnamefont {MacDonald}},\ }\href {https://doi.org/10.1038/s41563-020-00840-0} {\bibfield  {journal} {\bibinfo  {journal} {Nature materials}\ }\textbf {\bibinfo {volume} {19}},\ \bibinfo {pages} {1265} (\bibinfo {year} {2020})}\BibitemShut {NoStop}%
\bibitem [{\citenamefont {Angeli}\ and\ \citenamefont {MacDonald}(2021)}]{Angeli2021}%
  \BibitemOpen
  \bibfield  {author} {\bibinfo {author} {\bibfnamefont {M.}~\bibnamefont {Angeli}}\ and\ \bibinfo {author} {\bibfnamefont {A.~H.}\ \bibnamefont {MacDonald}},\ }\href {https://doi.org/10.1073/pnas.2021826118} {\bibfield  {journal} {\bibinfo  {journal} {Proceedings of the National Academy of Sciences}\ }\textbf {\bibinfo {volume} {118}},\ \bibinfo {pages} {e2021826118} (\bibinfo {year} {2021})}\BibitemShut {NoStop}%
\bibitem [{\citenamefont {Pan}\ \emph {et~al.}(2023)\citenamefont {Pan}, \citenamefont {Kim},\ and\ \citenamefont {Jian}}]{Pan2023}%
  \BibitemOpen
  \bibfield  {author} {\bibinfo {author} {\bibfnamefont {H.}~\bibnamefont {Pan}}, \bibinfo {author} {\bibfnamefont {E.-A.}\ \bibnamefont {Kim}}, \ and\ \bibinfo {author} {\bibfnamefont {C.-M.}\ \bibnamefont {Jian}},\ }\href {https://doi.org/10.1103/PhysRevResearch.5.043173} {\bibfield  {journal} {\bibinfo  {journal} {Physical Review Research}\ }\textbf {\bibinfo {volume} {5}},\ \bibinfo {pages} {043173} (\bibinfo {year} {2023})}\BibitemShut {NoStop}%
\bibitem [{\citenamefont {Foutty}\ \emph {et~al.}(2023)\citenamefont {Foutty}, \citenamefont {Yu}, \citenamefont {Devakul}, \citenamefont {Kometter}, \citenamefont {Zhang}, \citenamefont {Watanabe}, \citenamefont {Taniguchi}, \citenamefont {Fu},\ and\ \citenamefont {Feldman}}]{Foutty2023}%
  \BibitemOpen
  \bibfield  {author} {\bibinfo {author} {\bibfnamefont {B.~A.}\ \bibnamefont {Foutty}}, \bibinfo {author} {\bibfnamefont {J.}~\bibnamefont {Yu}}, \bibinfo {author} {\bibfnamefont {T.}~\bibnamefont {Devakul}}, \bibinfo {author} {\bibfnamefont {C.~R.}\ \bibnamefont {Kometter}}, \bibinfo {author} {\bibfnamefont {Y.}~\bibnamefont {Zhang}}, \bibinfo {author} {\bibfnamefont {K.}~\bibnamefont {Watanabe}}, \bibinfo {author} {\bibfnamefont {T.}~\bibnamefont {Taniguchi}}, \bibinfo {author} {\bibfnamefont {L.}~\bibnamefont {Fu}}, \ and\ \bibinfo {author} {\bibfnamefont {B.~E.}\ \bibnamefont {Feldman}},\ }\href {https://doi.org/10.1038/s41563-023-01534-z} {\bibfield  {journal} {\bibinfo  {journal} {Nature Materials}\ }\textbf {\bibinfo {volume} {22}},\ \bibinfo {pages} {731} (\bibinfo {year} {2023})}\BibitemShut {NoStop}%
\bibitem [{\citenamefont {Ma}\ \emph {et~al.}(2024)\citenamefont {Ma}, \citenamefont {Chaturvedi}, \citenamefont {Nguyen}, \citenamefont {Watanabe}, \citenamefont {Taniguchi}, \citenamefont {Mak},\ and\ \citenamefont {Shan}}]{Ma2024}%
  \BibitemOpen
  \bibfield  {author} {\bibinfo {author} {\bibfnamefont {L.}~\bibnamefont {Ma}}, \bibinfo {author} {\bibfnamefont {R.}~\bibnamefont {Chaturvedi}}, \bibinfo {author} {\bibfnamefont {P.~X.}\ \bibnamefont {Nguyen}}, \bibinfo {author} {\bibfnamefont {K.}~\bibnamefont {Watanabe}}, \bibinfo {author} {\bibfnamefont {T.}~\bibnamefont {Taniguchi}}, \bibinfo {author} {\bibfnamefont {K.~F.}\ \bibnamefont {Mak}}, \ and\ \bibinfo {author} {\bibfnamefont {J.}~\bibnamefont {Shan}},\ }\href {https://arxiv.org/abs/2412.07150} {\bibfield  {journal} {\bibinfo  {journal} {arXiv preprint arXiv:2412.07150}\ } (\bibinfo {year} {2024})}\BibitemShut {NoStop}%
\bibitem [{\citenamefont {Tolosa-Sime{\'o}n}\ \emph {et~al.}(2025)\citenamefont {Tolosa-Sime{\'o}n}, \citenamefont {Classen},\ and\ \citenamefont {Scherer}}]{Tolosa2025}%
  \BibitemOpen
  \bibfield  {author} {\bibinfo {author} {\bibfnamefont {M.}~\bibnamefont {Tolosa-Sime{\'o}n}}, \bibinfo {author} {\bibfnamefont {L.}~\bibnamefont {Classen}}, \ and\ \bibinfo {author} {\bibfnamefont {M.~M.}\ \bibnamefont {Scherer}},\ }\href {https://doi.org/10.1103/7kw4-8r3m} {\bibfield  {journal} {\bibinfo  {journal} {Physical Review B}\ }\textbf {\bibinfo {volume} {112}},\ \bibinfo {pages} {115133} (\bibinfo {year} {2025})}\BibitemShut {NoStop}%
\bibitem [{\citenamefont {Hasan}\ and\ \citenamefont {Kane}(2010)}]{Hasan2010}%
  \BibitemOpen
  \bibfield  {author} {\bibinfo {author} {\bibfnamefont {M.~Z.}\ \bibnamefont {Hasan}}\ and\ \bibinfo {author} {\bibfnamefont {C.~L.}\ \bibnamefont {Kane}},\ }\href {https://doi.org/10.1103/RevModPhys.82.3045} {\bibfield  {journal} {\bibinfo  {journal} {Reviews of Modern Physics}\ }\textbf {\bibinfo {volume} {82}},\ \bibinfo {pages} {3045} (\bibinfo {year} {2010})}\BibitemShut {NoStop}%
\bibitem [{\citenamefont {Qi}\ and\ \citenamefont {Zhang}(2011)}]{Qi2011}%
  \BibitemOpen
  \bibfield  {author} {\bibinfo {author} {\bibfnamefont {X.-L.}\ \bibnamefont {Qi}}\ and\ \bibinfo {author} {\bibfnamefont {S.-C.}\ \bibnamefont {Zhang}},\ }\href {https://doi.org/10.1103/RevModPhys.83.1057} {\bibfield  {journal} {\bibinfo  {journal} {Reviews of Modern Physics}\ }\textbf {\bibinfo {volume} {83}},\ \bibinfo {pages} {1057} (\bibinfo {year} {2011})}\BibitemShut {NoStop}%
\bibitem [{\citenamefont {Kozii}\ \emph {et~al.}(2019)\citenamefont {Kozii}, \citenamefont {Bi},\ and\ \citenamefont {Ruhman}}]{Kozii2019}%
  \BibitemOpen
  \bibfield  {author} {\bibinfo {author} {\bibfnamefont {V.}~\bibnamefont {Kozii}}, \bibinfo {author} {\bibfnamefont {Z.}~\bibnamefont {Bi}}, \ and\ \bibinfo {author} {\bibfnamefont {J.}~\bibnamefont {Ruhman}},\ }\href {https://doi.org/10.1103/PhysRevX.9.031046} {\bibfield  {journal} {\bibinfo  {journal} {Physical Review X}\ }\textbf {\bibinfo {volume} {9}},\ \bibinfo {pages} {031046} (\bibinfo {year} {2019})}\BibitemShut {NoStop}%
\bibitem [{\citenamefont {Kozii}\ \emph {et~al.}(2022)\citenamefont {Kozii}, \citenamefont {Klein}, \citenamefont {Fernandes},\ and\ \citenamefont {Ruhman}}]{Kozii2022}%
  \BibitemOpen
  \bibfield  {author} {\bibinfo {author} {\bibfnamefont {V.}~\bibnamefont {Kozii}}, \bibinfo {author} {\bibfnamefont {A.}~\bibnamefont {Klein}}, \bibinfo {author} {\bibfnamefont {R.~M.}\ \bibnamefont {Fernandes}}, \ and\ \bibinfo {author} {\bibfnamefont {J.}~\bibnamefont {Ruhman}},\ }\href {https://doi.org/10.1103/PhysRevLett.129.237001} {\bibfield  {journal} {\bibinfo  {journal} {Physical Review Letters}\ }\textbf {\bibinfo {volume} {129}},\ \bibinfo {pages} {237001} (\bibinfo {year} {2022})}\BibitemShut {NoStop}%
\bibitem [{\citenamefont {Montambaux}\ \emph {et~al.}(2009)\citenamefont {Montambaux}, \citenamefont {Pi\'echon}, \citenamefont {Fuchs},\ and\ \citenamefont {Goerbig}}]{Montambaux2009}%
  \BibitemOpen
  \bibfield  {author} {\bibinfo {author} {\bibfnamefont {G.}~\bibnamefont {Montambaux}}, \bibinfo {author} {\bibfnamefont {F.}~\bibnamefont {Pi\'echon}}, \bibinfo {author} {\bibfnamefont {J.-N.}\ \bibnamefont {Fuchs}}, \ and\ \bibinfo {author} {\bibfnamefont {M.~O.}\ \bibnamefont {Goerbig}},\ }\href {\doibase 10.1103/PhysRevB.80.153412} {\bibfield  {journal} {\bibinfo  {journal} {Phys. Rev. B}\ }\textbf {\bibinfo {volume} {80}},\ \bibinfo {pages} {153412} (\bibinfo {year} {2009})}\BibitemShut {NoStop}%
\bibitem [{\citenamefont {Isobe}\ \emph {et~al.}(2016)\citenamefont {Isobe}, \citenamefont {Yang}, \citenamefont {Chubukov}, \citenamefont {Schmalian},\ and\ \citenamefont {Nagaosa}}]{Isobe2015}%
  \BibitemOpen
  \bibfield  {author} {\bibinfo {author} {\bibfnamefont {H.}~\bibnamefont {Isobe}}, \bibinfo {author} {\bibfnamefont {B.-J.}\ \bibnamefont {Yang}}, \bibinfo {author} {\bibfnamefont {A.}~\bibnamefont {Chubukov}}, \bibinfo {author} {\bibfnamefont {J.}~\bibnamefont {Schmalian}}, \ and\ \bibinfo {author} {\bibfnamefont {N.}~\bibnamefont {Nagaosa}},\ }\href {\doibase 10.1103/PhysRevLett.116.076803} {\bibfield  {journal} {\bibinfo  {journal} {Phys. Rev. Lett.}\ }\textbf {\bibinfo {volume} {116}},\ \bibinfo {pages} {076803} (\bibinfo {year} {2016})}\BibitemShut {NoStop}%
\bibitem [{\citenamefont {Cho}\ and\ \citenamefont {Moon}(2016)}]{Cho2016novel}%
  \BibitemOpen
  \bibfield  {author} {\bibinfo {author} {\bibfnamefont {G.~Y.}\ \bibnamefont {Cho}}\ and\ \bibinfo {author} {\bibfnamefont {E.-G.}\ \bibnamefont {Moon}},\ }\href {https://doi.org/10.1038/srep19198} {\bibfield  {journal} {\bibinfo  {journal} {Scientific Reports}\ }\textbf {\bibinfo {volume} {6}},\ \bibinfo {pages} {19198} (\bibinfo {year} {2016})}\BibitemShut {NoStop}%
\bibitem [{\citenamefont {Link}\ \emph {et~al.}(2018)\citenamefont {Link}, \citenamefont {Narozhny}, \citenamefont {Kiselev},\ and\ \citenamefont {Schmalian}}]{Link2018}%
  \BibitemOpen
  \bibfield  {author} {\bibinfo {author} {\bibfnamefont {J.~M.}\ \bibnamefont {Link}}, \bibinfo {author} {\bibfnamefont {B.~N.}\ \bibnamefont {Narozhny}}, \bibinfo {author} {\bibfnamefont {E.~I.}\ \bibnamefont {Kiselev}}, \ and\ \bibinfo {author} {\bibfnamefont {J.}~\bibnamefont {Schmalian}},\ }\href {\doibase 10.1103/PhysRevLett.120.196801} {\bibfield  {journal} {\bibinfo  {journal} {Phys. Rev. Lett.}\ }\textbf {\bibinfo {volume} {120}},\ \bibinfo {pages} {196801} (\bibinfo {year} {2018})}\BibitemShut {NoStop}%
\bibitem [{\citenamefont {Gross}\ and\ \citenamefont {Neveu}(1974)}]{Gross1974}%
  \BibitemOpen
  \bibfield  {author} {\bibinfo {author} {\bibfnamefont {D.~J.}\ \bibnamefont {Gross}}\ and\ \bibinfo {author} {\bibfnamefont {A.}~\bibnamefont {Neveu}},\ }\href {https://doi.org/10.1103/PhysRevD.10.3235} {\bibfield  {journal} {\bibinfo  {journal} {Physical Review D}\ }\textbf {\bibinfo {volume} {10}},\ \bibinfo {pages} {3235} (\bibinfo {year} {1974})}\BibitemShut {NoStop}%
\bibitem [{\citenamefont {Zinn-Justin}(1991)}]{ZinnJustin1991}%
  \BibitemOpen
  \bibfield  {author} {\bibinfo {author} {\bibfnamefont {J.}~\bibnamefont {Zinn-Justin}},\ }\href {https://doi.org/10.1016/0550-3213(91)90043-W} {\bibfield  {journal} {\bibinfo  {journal} {Nuclear Physics B}\ }\textbf {\bibinfo {volume} {367}},\ \bibinfo {pages} {105} (\bibinfo {year} {1991})}\BibitemShut {NoStop}%
\bibitem [{\citenamefont {Herbut}(2006)}]{Herbut2006}%
  \BibitemOpen
  \bibfield  {author} {\bibinfo {author} {\bibfnamefont {I.~F.}\ \bibnamefont {Herbut}},\ }\href {https://doi.org/10.1103/PhysRevLett.97.146401} {\bibfield  {journal} {\bibinfo  {journal} {Physical Review Letters}\ }\textbf {\bibinfo {volume} {97}},\ \bibinfo {pages} {146401} (\bibinfo {year} {2006})}\BibitemShut {NoStop}%
\bibitem [{\citenamefont {Herbut}\ \emph {et~al.}(2009{\natexlab{a}})\citenamefont {Herbut}, \citenamefont {Juri\ifmmode \check{c}\else \v{c}\fi{}i\ifmmode~\acute{c}\else \'{c}\fi{}},\ and\ \citenamefont {Roy}}]{Herbut2009}%
  \BibitemOpen
  \bibfield  {author} {\bibinfo {author} {\bibfnamefont {I.~F.}\ \bibnamefont {Herbut}}, \bibinfo {author} {\bibfnamefont {V.}~\bibnamefont {Juri\ifmmode \check{c}\else \v{c}\fi{}i\ifmmode~\acute{c}\else \'{c}\fi{}}}, \ and\ \bibinfo {author} {\bibfnamefont {B.}~\bibnamefont {Roy}},\ }\href {\doibase 10.1103/PhysRevB.79.085116} {\bibfield  {journal} {\bibinfo  {journal} {Phys. Rev. B}\ }\textbf {\bibinfo {volume} {79}},\ \bibinfo {pages} {085116} (\bibinfo {year} {2009}{\natexlab{a}})}\BibitemShut {NoStop}%
\bibitem [{\citenamefont {Herbut}\ \emph {et~al.}(2009{\natexlab{b}})\citenamefont {Herbut}, \citenamefont {Juri\ifmmode \check{c}\else \v{c}\fi{}i\ifmmode~\acute{c}\else \'{c}\fi{}},\ and\ \citenamefont {Vafek}}]{Herbut2009b}%
  \BibitemOpen
  \bibfield  {author} {\bibinfo {author} {\bibfnamefont {I.~F.}\ \bibnamefont {Herbut}}, \bibinfo {author} {\bibfnamefont {V.}~\bibnamefont {Juri\ifmmode \check{c}\else \v{c}\fi{}i\ifmmode~\acute{c}\else \'{c}\fi{}}}, \ and\ \bibinfo {author} {\bibfnamefont {O.}~\bibnamefont {Vafek}},\ }\href {\doibase 10.1103/PhysRevB.80.075432} {\bibfield  {journal} {\bibinfo  {journal} {Phys. Rev. B}\ }\textbf {\bibinfo {volume} {80}},\ \bibinfo {pages} {075432} (\bibinfo {year} {2009}{\natexlab{b}})}\BibitemShut {NoStop}%
\bibitem [{\citenamefont {Juri\ifmmode \check{c}\else \v{c}\fi{}i\ifmmode~\acute{c}\else \'{c}\fi{}}\ \emph {et~al.}(2009)\citenamefont {Juri\ifmmode \check{c}\else \v{c}\fi{}i\ifmmode~\acute{c}\else \'{c}\fi{}}, \citenamefont {Herbut},\ and\ \citenamefont {Semenoff}}]{Jurivcic2009}%
  \BibitemOpen
  \bibfield  {author} {\bibinfo {author} {\bibfnamefont {V.}~\bibnamefont {Juri\ifmmode \check{c}\else \v{c}\fi{}i\ifmmode~\acute{c}\else \'{c}\fi{}}}, \bibinfo {author} {\bibfnamefont {I.~F.}\ \bibnamefont {Herbut}}, \ and\ \bibinfo {author} {\bibfnamefont {G.~W.}\ \bibnamefont {Semenoff}},\ }\href {\doibase 10.1103/PhysRevB.80.081405} {\bibfield  {journal} {\bibinfo  {journal} {Phys. Rev. B}\ }\textbf {\bibinfo {volume} {80}},\ \bibinfo {pages} {081405} (\bibinfo {year} {2009})}\BibitemShut {NoStop}%
\bibitem [{\citenamefont {Weeks}\ and\ \citenamefont {Franz}(2010)}]{Weeks2010}%
  \BibitemOpen
  \bibfield  {author} {\bibinfo {author} {\bibfnamefont {C.}~\bibnamefont {Weeks}}\ and\ \bibinfo {author} {\bibfnamefont {M.}~\bibnamefont {Franz}},\ }\href {\doibase 10.1103/PhysRevB.81.085105} {\bibfield  {journal} {\bibinfo  {journal} {Phys. Rev. B}\ }\textbf {\bibinfo {volume} {81}},\ \bibinfo {pages} {085105} (\bibinfo {year} {2010})}\BibitemShut {NoStop}%
\bibitem [{\citenamefont {Semenoff}(2012)}]{Semenoff2012}%
  \BibitemOpen
  \bibfield  {author} {\bibinfo {author} {\bibfnamefont {G.~W.}\ \bibnamefont {Semenoff}},\ }\href {https://dx.doi.org/10.1088/0031-8949/2012/T146/014016} {\bibfield  {journal} {\bibinfo  {journal} {Physica Scripta}\ }\textbf {\bibinfo {volume} {2012}},\ \bibinfo {pages} {014016} (\bibinfo {year} {2012})}\BibitemShut {NoStop}%
\bibitem [{\citenamefont {Assaad}\ and\ \citenamefont {Herbut}(2013)}]{Assaad2013}%
  \BibitemOpen
  \bibfield  {author} {\bibinfo {author} {\bibfnamefont {F.~F.}\ \bibnamefont {Assaad}}\ and\ \bibinfo {author} {\bibfnamefont {I.~F.}\ \bibnamefont {Herbut}},\ }\href {https://doi.org/10.1103/PhysRevX.3.031010} {\bibfield  {journal} {\bibinfo  {journal} {Physical Review X}\ }\textbf {\bibinfo {volume} {3}},\ \bibinfo {pages} {031010} (\bibinfo {year} {2013})}\BibitemShut {NoStop}%
\bibitem [{\citenamefont {Han}\ and\ \citenamefont {Moon}(2018)}]{Han2018}%
  \BibitemOpen
  \bibfield  {author} {\bibinfo {author} {\bibfnamefont {S.}~\bibnamefont {Han}}\ and\ \bibinfo {author} {\bibfnamefont {E.-G.}\ \bibnamefont {Moon}},\ }\href {\doibase 10.1103/PhysRevB.97.241101} {\bibfield  {journal} {\bibinfo  {journal} {Phys. Rev. B}\ }\textbf {\bibinfo {volume} {97}},\ \bibinfo {pages} {241101} (\bibinfo {year} {2018})}\BibitemShut {NoStop}%
\bibitem [{\citenamefont {Ihrig}\ \emph {et~al.}(2018)\citenamefont {Ihrig}, \citenamefont {Mihaila},\ and\ \citenamefont {Scherer}}]{Ihrig2018}%
  \BibitemOpen
  \bibfield  {author} {\bibinfo {author} {\bibfnamefont {B.}~\bibnamefont {Ihrig}}, \bibinfo {author} {\bibfnamefont {L.~N.}\ \bibnamefont {Mihaila}}, \ and\ \bibinfo {author} {\bibfnamefont {M.~M.}\ \bibnamefont {Scherer}},\ }\href {\doibase 10.1103/PhysRevB.98.125109} {\bibfield  {journal} {\bibinfo  {journal} {Phys. Rev. B}\ }\textbf {\bibinfo {volume} {98}},\ \bibinfo {pages} {125109} (\bibinfo {year} {2018})}\BibitemShut {NoStop}%
\bibitem [{\citenamefont {Lang}\ and\ \citenamefont {L\"auchli}(2019)}]{Lang2019}%
  \BibitemOpen
  \bibfield  {author} {\bibinfo {author} {\bibfnamefont {T.~C.}\ \bibnamefont {Lang}}\ and\ \bibinfo {author} {\bibfnamefont {A.~M.}\ \bibnamefont {L\"auchli}},\ }\href {\doibase 10.1103/PhysRevLett.123.137602} {\bibfield  {journal} {\bibinfo  {journal} {Phys. Rev. Lett.}\ }\textbf {\bibinfo {volume} {123}},\ \bibinfo {pages} {137602} (\bibinfo {year} {2019})}\BibitemShut {NoStop}%
\bibitem [{\citenamefont {Parthenios}\ and\ \citenamefont {Classen}(2023)}]{Parthenios2023}%
  \BibitemOpen
  \bibfield  {author} {\bibinfo {author} {\bibfnamefont {N.}~\bibnamefont {Parthenios}}\ and\ \bibinfo {author} {\bibfnamefont {L.}~\bibnamefont {Classen}},\ }\href {\doibase 10.1103/PhysRevB.108.235120} {\bibfield  {journal} {\bibinfo  {journal} {Phys. Rev. B}\ }\textbf {\bibinfo {volume} {108}},\ \bibinfo {pages} {235120} (\bibinfo {year} {2023})}\BibitemShut {NoStop}%
\bibitem [{\citenamefont {Biedermann}\ and\ \citenamefont {Janssen}(2025)}]{Biedermann2025}%
  \BibitemOpen
  \bibfield  {author} {\bibinfo {author} {\bibfnamefont {J.}~\bibnamefont {Biedermann}}\ and\ \bibinfo {author} {\bibfnamefont {L.}~\bibnamefont {Janssen}},\ }\href {\doibase 10.1103/hj61-dw78} {\bibfield  {journal} {\bibinfo  {journal} {Phys. Rev. B}\ }\textbf {\bibinfo {volume} {112}},\ \bibinfo {pages} {L041109} (\bibinfo {year} {2025})}\BibitemShut {NoStop}%
\bibitem [{\citenamefont {Hawashin}\ \emph {et~al.}(2025)\citenamefont {Hawashin}, \citenamefont {Scherer},\ and\ \citenamefont {Janssen}}]{Hawashin2025}%
  \BibitemOpen
  \bibfield  {author} {\bibinfo {author} {\bibfnamefont {B.}~\bibnamefont {Hawashin}}, \bibinfo {author} {\bibfnamefont {M.~M.}\ \bibnamefont {Scherer}}, \ and\ \bibinfo {author} {\bibfnamefont {L.}~\bibnamefont {Janssen}},\ }\href {\doibase 10.1103/PhysRevB.111.205129} {\bibfield  {journal} {\bibinfo  {journal} {Phys. Rev. B}\ }\textbf {\bibinfo {volume} {111}},\ \bibinfo {pages} {205129} (\bibinfo {year} {2025})}\BibitemShut {NoStop}%
\bibitem [{\citenamefont {Huang}\ \emph {et~al.}(2025)\citenamefont {Huang}, \citenamefont {Parthenios}, \citenamefont {Ulybyshev}, \citenamefont {Zhang}, \citenamefont {Assaad}, \citenamefont {Classen},\ and\ \citenamefont {Meng}}]{Huang2025}%
  \BibitemOpen
  \bibfield  {author} {\bibinfo {author} {\bibfnamefont {C.}~\bibnamefont {Huang}}, \bibinfo {author} {\bibfnamefont {N.}~\bibnamefont {Parthenios}}, \bibinfo {author} {\bibfnamefont {M.}~\bibnamefont {Ulybyshev}}, \bibinfo {author} {\bibfnamefont {X.}~\bibnamefont {Zhang}}, \bibinfo {author} {\bibfnamefont {F.~F.}\ \bibnamefont {Assaad}}, \bibinfo {author} {\bibfnamefont {L.}~\bibnamefont {Classen}}, \ and\ \bibinfo {author} {\bibfnamefont {Z.~Y.}\ \bibnamefont {Meng}},\ }\href {https://doi.org/10.1038/s41467-025-62461-y} {\bibfield  {journal} {\bibinfo  {journal} {Nature Communications}\ }\textbf {\bibinfo {volume} {16}},\ \bibinfo {pages} {7176} (\bibinfo {year} {2025})}\BibitemShut {NoStop}%
\bibitem [{\citenamefont {Lang}\ and\ \citenamefont {L{\"a}uchli}(2025)}]{Lang2025}%
  \BibitemOpen
  \bibfield  {author} {\bibinfo {author} {\bibfnamefont {T.~C.}\ \bibnamefont {Lang}}\ and\ \bibinfo {author} {\bibfnamefont {A.~M.}\ \bibnamefont {L{\"a}uchli}},\ }\href {https://arxiv.org/abs/2503.15000} {\bibfield  {journal} {\bibinfo  {journal} {arXiv preprint arXiv:2503.15000}\ } (\bibinfo {year} {2025})}\BibitemShut {NoStop}%
\bibitem [{\citenamefont {Bultinck}\ \emph {et~al.}(2020)\citenamefont {Bultinck}, \citenamefont {Khalaf}, \citenamefont {Liu}, \citenamefont {Chatterjee}, \citenamefont {Vishwanath},\ and\ \citenamefont {Zaletel}}]{Bultinck2020}%
  \BibitemOpen
  \bibfield  {author} {\bibinfo {author} {\bibfnamefont {N.}~\bibnamefont {Bultinck}}, \bibinfo {author} {\bibfnamefont {E.}~\bibnamefont {Khalaf}}, \bibinfo {author} {\bibfnamefont {S.}~\bibnamefont {Liu}}, \bibinfo {author} {\bibfnamefont {S.}~\bibnamefont {Chatterjee}}, \bibinfo {author} {\bibfnamefont {A.}~\bibnamefont {Vishwanath}}, \ and\ \bibinfo {author} {\bibfnamefont {M.~P.}\ \bibnamefont {Zaletel}},\ }\href {\doibase 10.1103/PhysRevX.10.031034} {\bibfield  {journal} {\bibinfo  {journal} {Phys. Rev. X}\ }\textbf {\bibinfo {volume} {10}},\ \bibinfo {pages} {031034} (\bibinfo {year} {2020})}\BibitemShut {NoStop}%
\bibitem [{\citenamefont {Liu}\ \emph {et~al.}(2021)\citenamefont {Liu}, \citenamefont {Khalaf}, \citenamefont {Lee},\ and\ \citenamefont {Vishwanath}}]{Liu2021}%
  \BibitemOpen
  \bibfield  {author} {\bibinfo {author} {\bibfnamefont {S.}~\bibnamefont {Liu}}, \bibinfo {author} {\bibfnamefont {E.}~\bibnamefont {Khalaf}}, \bibinfo {author} {\bibfnamefont {J.~Y.}\ \bibnamefont {Lee}}, \ and\ \bibinfo {author} {\bibfnamefont {A.}~\bibnamefont {Vishwanath}},\ }\href {\doibase 10.1103/PhysRevResearch.3.013033} {\bibfield  {journal} {\bibinfo  {journal} {Phys. Rev. Res.}\ }\textbf {\bibinfo {volume} {3}},\ \bibinfo {pages} {013033} (\bibinfo {year} {2021})}\BibitemShut {NoStop}%
\bibitem [{\citenamefont {Ledwith}\ \emph {et~al.}(2021)\citenamefont {Ledwith}, \citenamefont {Khalaf},\ and\ \citenamefont {Vishwanath}}]{Ledwith2021strong}%
  \BibitemOpen
  \bibfield  {author} {\bibinfo {author} {\bibfnamefont {P.~J.}\ \bibnamefont {Ledwith}}, \bibinfo {author} {\bibfnamefont {E.}~\bibnamefont {Khalaf}}, \ and\ \bibinfo {author} {\bibfnamefont {A.}~\bibnamefont {Vishwanath}},\ }\href {https://doi.org/10.1016/j.aop.2021.168646} {\bibfield  {journal} {\bibinfo  {journal} {Annals of Physics}\ }\textbf {\bibinfo {volume} {435}},\ \bibinfo {pages} {168646} (\bibinfo {year} {2021})}\BibitemShut {NoStop}%
\bibitem [{\citenamefont {Nuckolls}\ \emph {et~al.}(2023)\citenamefont {Nuckolls}, \citenamefont {Lee}, \citenamefont {Oh}, \citenamefont {Wong}, \citenamefont {Soejima}, \citenamefont {Hong}, \citenamefont {C{\u{a}}lug{\u{a}}ru}, \citenamefont {Herzog-Arbeitman}, \citenamefont {Bernevig}, \citenamefont {Watanabe} \emph {et~al.}}]{Nuckolls2023}%
  \BibitemOpen
  \bibfield  {author} {\bibinfo {author} {\bibfnamefont {K.~P.}\ \bibnamefont {Nuckolls}}, \bibinfo {author} {\bibfnamefont {R.~L.}\ \bibnamefont {Lee}}, \bibinfo {author} {\bibfnamefont {M.}~\bibnamefont {Oh}}, \bibinfo {author} {\bibfnamefont {D.}~\bibnamefont {Wong}}, \bibinfo {author} {\bibfnamefont {T.}~\bibnamefont {Soejima}}, \bibinfo {author} {\bibfnamefont {J.~P.}\ \bibnamefont {Hong}}, \bibinfo {author} {\bibfnamefont {D.}~\bibnamefont {C{\u{a}}lug{\u{a}}ru}}, \bibinfo {author} {\bibfnamefont {J.}~\bibnamefont {Herzog-Arbeitman}}, \bibinfo {author} {\bibfnamefont {B.~A.}\ \bibnamefont {Bernevig}}, \bibinfo {author} {\bibfnamefont {K.}~\bibnamefont {Watanabe}},  \emph {et~al.},\ }\href {https://doi.org/10.1038/s41586-023-06226-x} {\bibfield  {journal} {\bibinfo  {journal} {Nature}\ }\textbf {\bibinfo {volume} {620}},\ \bibinfo {pages} {525} (\bibinfo {year} {2023})}\BibitemShut {NoStop}%
\bibitem [{\citenamefont {Chandrasekharan}\ and\ \citenamefont {Li}(2013)}]{Chandrasekharan2013}%
  \BibitemOpen
  \bibfield  {author} {\bibinfo {author} {\bibfnamefont {S.}~\bibnamefont {Chandrasekharan}}\ and\ \bibinfo {author} {\bibfnamefont {A.}~\bibnamefont {Li}},\ }\href {\doibase 10.1103/PhysRevD.88.021701} {\bibfield  {journal} {\bibinfo  {journal} {Phys. Rev. D}\ }\textbf {\bibinfo {volume} {88}},\ \bibinfo {pages} {021701} (\bibinfo {year} {2013})}\BibitemShut {NoStop}%
\bibitem [{\citenamefont {Iliesiu}\ \emph {et~al.}(2018)\citenamefont {Iliesiu}, \citenamefont {Kos}, \citenamefont {Poland}, \citenamefont {Pufu},\ and\ \citenamefont {Simmons-Duffin}}]{Iliesiu2018}%
  \BibitemOpen
  \bibfield  {author} {\bibinfo {author} {\bibfnamefont {L.}~\bibnamefont {Iliesiu}}, \bibinfo {author} {\bibfnamefont {F.}~\bibnamefont {Kos}}, \bibinfo {author} {\bibfnamefont {D.}~\bibnamefont {Poland}}, \bibinfo {author} {\bibfnamefont {S.~S.}\ \bibnamefont {Pufu}}, \ and\ \bibinfo {author} {\bibfnamefont {D.}~\bibnamefont {Simmons-Duffin}},\ }\href {https://doi.org/10.1007/JHEP01(2018)036} {\bibfield  {journal} {\bibinfo  {journal} {Journal of High Energy Physics}\ }\textbf {\bibinfo {volume} {36}},\ \bibinfo {pages} {1} (\bibinfo {year} {2018})}\BibitemShut {NoStop}%
\bibitem [{\citenamefont {Vacca}\ and\ \citenamefont {Zambelli}(2015)}]{Vacca2015}%
  \BibitemOpen
  \bibfield  {author} {\bibinfo {author} {\bibfnamefont {G.~P.}\ \bibnamefont {Vacca}}\ and\ \bibinfo {author} {\bibfnamefont {L.}~\bibnamefont {Zambelli}},\ }\href@noop {} {\bibfield  {journal} {\bibinfo  {journal} {Physical Review D}\ }\textbf {\bibinfo {volume} {91}},\ \bibinfo {pages} {125003} (\bibinfo {year} {2015})}\BibitemShut {NoStop}%
\bibitem [{\citenamefont {Knorr}(2016)}]{Knorr2016}%
  \BibitemOpen
  \bibfield  {author} {\bibinfo {author} {\bibfnamefont {B.}~\bibnamefont {Knorr}},\ }\href {\doibase 10.1103/PhysRevB.94.245102} {\bibfield  {journal} {\bibinfo  {journal} {Phys. Rev. B}\ }\textbf {\bibinfo {volume} {94}},\ \bibinfo {pages} {245102} (\bibinfo {year} {2016})}\BibitemShut {NoStop}%
\bibitem [{\citenamefont {Sachdev}\ and\ \citenamefont {Ye}(1993)}]{sachdev1993}%
  \BibitemOpen
  \bibfield  {author} {\bibinfo {author} {\bibfnamefont {S.}~\bibnamefont {Sachdev}}\ and\ \bibinfo {author} {\bibfnamefont {J.}~\bibnamefont {Ye}},\ }\href {\doibase 10.1103/PhysRevLett.70.3339} {\bibfield  {journal} {\bibinfo  {journal} {Phys. Rev. Lett.}\ }\textbf {\bibinfo {volume} {70}},\ \bibinfo {pages} {3339} (\bibinfo {year} {1993})}\BibitemShut {NoStop}%
\bibitem [{\citenamefont {Georges}\ \emph {et~al.}(2000)\citenamefont {Georges}, \citenamefont {Parcollet},\ and\ \citenamefont {Sachdev}}]{georges2000}%
  \BibitemOpen
  \bibfield  {author} {\bibinfo {author} {\bibfnamefont {A.}~\bibnamefont {Georges}}, \bibinfo {author} {\bibfnamefont {O.}~\bibnamefont {Parcollet}}, \ and\ \bibinfo {author} {\bibfnamefont {S.}~\bibnamefont {Sachdev}},\ }\href {\doibase 10.1103/PhysRevLett.85.840} {\bibfield  {journal} {\bibinfo  {journal} {Phys. Rev. Lett.}\ }\textbf {\bibinfo {volume} {85}},\ \bibinfo {pages} {840} (\bibinfo {year} {2000})}\BibitemShut {NoStop}%
\bibitem [{\citenamefont {Sachdev}(2010)}]{sachdev2010}%
  \BibitemOpen
  \bibfield  {author} {\bibinfo {author} {\bibfnamefont {S.}~\bibnamefont {Sachdev}},\ }\href {\doibase 10.1103/PhysRevLett.105.151602} {\bibfield  {journal} {\bibinfo  {journal} {Phys. Rev. Lett.}\ }\textbf {\bibinfo {volume} {105}},\ \bibinfo {pages} {151602} (\bibinfo {year} {2010})}\BibitemShut {NoStop}%
\bibitem [{\citenamefont {Kitaev}(2015)}]{kitaev2015}%
  \BibitemOpen
  \bibfield  {author} {\bibinfo {author} {\bibfnamefont {A.}~\bibnamefont {Kitaev}},\ }\href {http://online.kitp.ucsb.edu/online/entangled15/} {\bibfield  {journal} {\bibinfo  {journal} {Talks at KITP, University of California, Santa Barbara, Entanglement in Strongly-Correlated Quantum Matter}\ } (\bibinfo {year} {2015})}\BibitemShut {NoStop}%
\bibitem [{\citenamefont {Esterlis}\ and\ \citenamefont {Schmalian}(2019)}]{Esterlis2019}%
  \BibitemOpen
  \bibfield  {author} {\bibinfo {author} {\bibfnamefont {I.}~\bibnamefont {Esterlis}}\ and\ \bibinfo {author} {\bibfnamefont {J.}~\bibnamefont {Schmalian}},\ }\href {\doibase 10.1103/PhysRevB.100.115132} {\bibfield  {journal} {\bibinfo  {journal} {Phys. Rev. B}\ }\textbf {\bibinfo {volume} {100}},\ \bibinfo {pages} {115132} (\bibinfo {year} {2019})}\BibitemShut {NoStop}%
\bibitem [{\citenamefont {Wang}(2020)}]{wang2020}%
  \BibitemOpen
  \bibfield  {author} {\bibinfo {author} {\bibfnamefont {Y.}~\bibnamefont {Wang}},\ }\href {\doibase 10.1103/PhysRevLett.124.017002} {\bibfield  {journal} {\bibinfo  {journal} {Phys. Rev. Lett.}\ }\textbf {\bibinfo {volume} {124}},\ \bibinfo {pages} {017002} (\bibinfo {year} {2020})}\BibitemShut {NoStop}%
\bibitem [{\citenamefont {Hauck}\ \emph {et~al.}(2020)\citenamefont {Hauck}, \citenamefont {Klug}, \citenamefont {Esterlis},\ and\ \citenamefont {Schmalian}}]{Hauck2020}%
  \BibitemOpen
  \bibfield  {author} {\bibinfo {author} {\bibfnamefont {D.}~\bibnamefont {Hauck}}, \bibinfo {author} {\bibfnamefont {M.~J.}\ \bibnamefont {Klug}}, \bibinfo {author} {\bibfnamefont {I.}~\bibnamefont {Esterlis}}, \ and\ \bibinfo {author} {\bibfnamefont {J.}~\bibnamefont {Schmalian}},\ }\href {https://doi.org/10.1016/j.aop.2020.168120} {\bibfield  {journal} {\bibinfo  {journal} {Annals of Physics}\ }\textbf {\bibinfo {volume} {417}},\ \bibinfo {pages} {168120} (\bibinfo {year} {2020})}\BibitemShut {NoStop}%
\bibitem [{\citenamefont {Esterlis}\ and\ \citenamefont {Schmalian}(2025)}]{Esterlis2025}%
  \BibitemOpen
  \bibfield  {author} {\bibinfo {author} {\bibfnamefont {I.}~\bibnamefont {Esterlis}}\ and\ \bibinfo {author} {\bibfnamefont {J.}~\bibnamefont {Schmalian}},\ }\href {https://arxiv.org/abs/2506.11952} {\bibfield  {journal} {\bibinfo  {journal} {arXiv preprint arXiv:2506.11952}\ } (\bibinfo {year} {2025})}\BibitemShut {NoStop}%
\bibitem [{\citenamefont {Maldacena}\ \emph {et~al.}(2016)\citenamefont {Maldacena}, \citenamefont {Shenker},\ and\ \citenamefont {Stanford}}]{Maldacena2016}%
  \BibitemOpen
  \bibfield  {author} {\bibinfo {author} {\bibfnamefont {J.}~\bibnamefont {Maldacena}}, \bibinfo {author} {\bibfnamefont {S.~H.}\ \bibnamefont {Shenker}}, \ and\ \bibinfo {author} {\bibfnamefont {D.}~\bibnamefont {Stanford}},\ }\href {http://dx.doi.org/10.1007/JHEP08(2016)106} {\bibfield  {journal} {\bibinfo  {journal} {Journal of High Energy Physics}\ }\textbf {\bibinfo {volume} {106}},\ \bibinfo {pages} {1} (\bibinfo {year} {2016})}\BibitemShut {NoStop}%
\bibitem [{\citenamefont {Chowdhury}\ \emph {et~al.}(2018)\citenamefont {Chowdhury}, \citenamefont {Werman}, \citenamefont {Berg},\ and\ \citenamefont {Senthil}}]{Chowdhury2018}%
  \BibitemOpen
  \bibfield  {author} {\bibinfo {author} {\bibfnamefont {D.}~\bibnamefont {Chowdhury}}, \bibinfo {author} {\bibfnamefont {Y.}~\bibnamefont {Werman}}, \bibinfo {author} {\bibfnamefont {E.}~\bibnamefont {Berg}}, \ and\ \bibinfo {author} {\bibfnamefont {T.}~\bibnamefont {Senthil}},\ }\href {\doibase 10.1103/PhysRevX.8.031024} {\bibfield  {journal} {\bibinfo  {journal} {Phys. Rev. X}\ }\textbf {\bibinfo {volume} {8}},\ \bibinfo {pages} {031024} (\bibinfo {year} {2018})}\BibitemShut {NoStop}%
\bibitem [{\citenamefont {Chowdhury}\ and\ \citenamefont {Berg}(2020)}]{Chowdhury2020}%
  \BibitemOpen
  \bibfield  {author} {\bibinfo {author} {\bibfnamefont {D.}~\bibnamefont {Chowdhury}}\ and\ \bibinfo {author} {\bibfnamefont {E.}~\bibnamefont {Berg}},\ }\href {\doibase 10.1103/PhysRevResearch.2.013301} {\bibfield  {journal} {\bibinfo  {journal} {Phys. Rev. Res.}\ }\textbf {\bibinfo {volume} {2}},\ \bibinfo {pages} {013301} (\bibinfo {year} {2020})}\BibitemShut {NoStop}%
\bibitem [{\citenamefont {Esterlis}\ \emph {et~al.}(2021)\citenamefont {Esterlis}, \citenamefont {Guo}, \citenamefont {Patel},\ and\ \citenamefont {Sachdev}}]{Esterlis2021}%
  \BibitemOpen
  \bibfield  {author} {\bibinfo {author} {\bibfnamefont {I.}~\bibnamefont {Esterlis}}, \bibinfo {author} {\bibfnamefont {H.}~\bibnamefont {Guo}}, \bibinfo {author} {\bibfnamefont {A.~A.}\ \bibnamefont {Patel}}, \ and\ \bibinfo {author} {\bibfnamefont {S.}~\bibnamefont {Sachdev}},\ }\href {\doibase 10.1103/PhysRevB.103.235129} {\bibfield  {journal} {\bibinfo  {journal} {Phys. Rev. B}\ }\textbf {\bibinfo {volume} {103}},\ \bibinfo {pages} {235129} (\bibinfo {year} {2021})}\BibitemShut {NoStop}%
\bibitem [{\citenamefont {Patel}\ \emph {et~al.}(2023)\citenamefont {Patel}, \citenamefont {Guo}, \citenamefont {Esterlis},\ and\ \citenamefont {Sachdev}}]{Patel2023}%
  \BibitemOpen
  \bibfield  {author} {\bibinfo {author} {\bibfnamefont {A.~A.}\ \bibnamefont {Patel}}, \bibinfo {author} {\bibfnamefont {H.}~\bibnamefont {Guo}}, \bibinfo {author} {\bibfnamefont {I.}~\bibnamefont {Esterlis}}, \ and\ \bibinfo {author} {\bibfnamefont {S.}~\bibnamefont {Sachdev}},\ }\href {https://doi.org/10.1126/science.abq6011} {\bibfield  {journal} {\bibinfo  {journal} {Science}\ }\textbf {\bibinfo {volume} {381}},\ \bibinfo {pages} {790} (\bibinfo {year} {2023})}\BibitemShut {NoStop}%
\bibitem [{\citenamefont {Li}\ \emph {et~al.}(2024)\citenamefont {Li}, \citenamefont {Valentinis}, \citenamefont {Patel}, \citenamefont {Guo}, \citenamefont {Schmalian}, \citenamefont {Sachdev},\ and\ \citenamefont {Esterlis}}]{Li2024}%
  \BibitemOpen
  \bibfield  {author} {\bibinfo {author} {\bibfnamefont {C.}~\bibnamefont {Li}}, \bibinfo {author} {\bibfnamefont {D.}~\bibnamefont {Valentinis}}, \bibinfo {author} {\bibfnamefont {A.~A.}\ \bibnamefont {Patel}}, \bibinfo {author} {\bibfnamefont {H.}~\bibnamefont {Guo}}, \bibinfo {author} {\bibfnamefont {J.}~\bibnamefont {Schmalian}}, \bibinfo {author} {\bibfnamefont {S.}~\bibnamefont {Sachdev}}, \ and\ \bibinfo {author} {\bibfnamefont {I.}~\bibnamefont {Esterlis}},\ }\href {\doibase 10.1103/PhysRevLett.133.186502} {\bibfield  {journal} {\bibinfo  {journal} {Phys. Rev. Lett.}\ }\textbf {\bibinfo {volume} {133}},\ \bibinfo {pages} {186502} (\bibinfo {year} {2024})}\BibitemShut {NoStop}%
\bibitem [{\citenamefont {Kopnin}\ and\ \citenamefont {Sonin}(2008)}]{Kopnin2008}%
  \BibitemOpen
  \bibfield  {author} {\bibinfo {author} {\bibfnamefont {N.~B.}\ \bibnamefont {Kopnin}}\ and\ \bibinfo {author} {\bibfnamefont {E.~B.}\ \bibnamefont {Sonin}},\ }\href {\doibase 10.1103/PhysRevLett.100.246808} {\bibfield  {journal} {\bibinfo  {journal} {Phys. Rev. Lett.}\ }\textbf {\bibinfo {volume} {100}},\ \bibinfo {pages} {246808} (\bibinfo {year} {2008})}\BibitemShut {NoStop}%
\bibitem [{\citenamefont {Abah}\ and\ \citenamefont {Kiselev}(2011)}]{Abah2011}%
  \BibitemOpen
  \bibfield  {author} {\bibinfo {author} {\bibfnamefont {O.}~\bibnamefont {Abah}}\ and\ \bibinfo {author} {\bibfnamefont {M.}~\bibnamefont {Kiselev}},\ }\href {https://doi.org/10.1140/epjb/e2011-10901-0} {\bibfield  {journal} {\bibinfo  {journal} {The Europ. Phys. Journ. B}\ }\textbf {\bibinfo {volume} {82}},\ \bibinfo {pages} {47} (\bibinfo {year} {2011})}\BibitemShut {NoStop}%
\bibitem [{\citenamefont {Wiegmann}(1999)}]{Wiegmann1999}%
  \BibitemOpen
  \bibfield  {author} {\bibinfo {author} {\bibfnamefont {P.~B.}\ \bibnamefont {Wiegmann}},\ }\href {\doibase 10.1103/PhysRevB.59.15705} {\bibfield  {journal} {\bibinfo  {journal} {Phys. Rev. B}\ }\textbf {\bibinfo {volume} {59}},\ \bibinfo {pages} {15705} (\bibinfo {year} {1999})}\BibitemShut {NoStop}%
\bibitem [{\citenamefont {Abanov}\ and\ \citenamefont {Wiegmann}(2000)}]{Abanov2000theta}%
  \BibitemOpen
  \bibfield  {author} {\bibinfo {author} {\bibfnamefont {A.}~\bibnamefont {Abanov}}\ and\ \bibinfo {author} {\bibfnamefont {P.~B.}\ \bibnamefont {Wiegmann}},\ }\href {https://doi.org/10.1016/S0550-3213(99)00820-2} {\bibfield  {journal} {\bibinfo  {journal} {Nuclear Physics B}\ }\textbf {\bibinfo {volume} {570}},\ \bibinfo {pages} {685} (\bibinfo {year} {2000})}\BibitemShut {NoStop}%
\bibitem [{\citenamefont {Grover}\ and\ \citenamefont {Senthil}(2008)}]{Grover2008}%
  \BibitemOpen
  \bibfield  {author} {\bibinfo {author} {\bibfnamefont {T.}~\bibnamefont {Grover}}\ and\ \bibinfo {author} {\bibfnamefont {T.}~\bibnamefont {Senthil}},\ }\href {\doibase 10.1103/PhysRevLett.100.156804} {\bibfield  {journal} {\bibinfo  {journal} {Phys. Rev. Lett.}\ }\textbf {\bibinfo {volume} {100}},\ \bibinfo {pages} {156804} (\bibinfo {year} {2008})}\BibitemShut {NoStop}%
\bibitem [{\citenamefont {Christos}\ \emph {et~al.}(2020)\citenamefont {Christos}, \citenamefont {Sachdev},\ and\ \citenamefont {Scheurer}}]{Christos2020}%
  \BibitemOpen
  \bibfield  {author} {\bibinfo {author} {\bibfnamefont {M.}~\bibnamefont {Christos}}, \bibinfo {author} {\bibfnamefont {S.}~\bibnamefont {Sachdev}}, \ and\ \bibinfo {author} {\bibfnamefont {M.~S.}\ \bibnamefont {Scheurer}},\ }\href {https://doi.org/10.1073/pnas.2014691117} {\bibfield  {journal} {\bibinfo  {journal} {Proceedings of the National Academy of Sciences}\ }\textbf {\bibinfo {volume} {117}},\ \bibinfo {pages} {29543} (\bibinfo {year} {2020})}\BibitemShut {NoStop}%
\bibitem [{\citenamefont {Khalaf}\ \emph {et~al.}(2021)\citenamefont {Khalaf}, \citenamefont {Chatterjee}, \citenamefont {Bultinck}, \citenamefont {Zaletel},\ and\ \citenamefont {Vishwanath}}]{Khalaf2021charged}%
  \BibitemOpen
  \bibfield  {author} {\bibinfo {author} {\bibfnamefont {E.}~\bibnamefont {Khalaf}}, \bibinfo {author} {\bibfnamefont {S.}~\bibnamefont {Chatterjee}}, \bibinfo {author} {\bibfnamefont {N.}~\bibnamefont {Bultinck}}, \bibinfo {author} {\bibfnamefont {M.~P.}\ \bibnamefont {Zaletel}}, \ and\ \bibinfo {author} {\bibfnamefont {A.}~\bibnamefont {Vishwanath}},\ }\href {https://doi.org/10.1126/sciadv.abf5299} {\bibfield  {journal} {\bibinfo  {journal} {Science Advances}\ }\textbf {\bibinfo {volume} {7}},\ \bibinfo {pages} {eabf5299} (\bibinfo {year} {2021})}\BibitemShut {NoStop}%
\bibitem [{\citenamefont {Stangier}\ \emph {et~al.}(2025)\citenamefont {Stangier}, \citenamefont {Scheurer}, \citenamefont {Sheehy},\ and\ \citenamefont {Schmalian}}]{Stangier2025}%
  \BibitemOpen
  \bibfield  {author} {\bibinfo {author} {\bibfnamefont {V.~C.}\ \bibnamefont {Stangier}}, \bibinfo {author} {\bibfnamefont {M.~S.}\ \bibnamefont {Scheurer}}, \bibinfo {author} {\bibfnamefont {D.~E.}\ \bibnamefont {Sheehy}}, \ and\ \bibinfo {author} {\bibfnamefont {J.}~\bibnamefont {Schmalian}},\ }\href {https://arxiv.org/abs/2510.06313} {\bibfield  {journal} {\bibinfo  {journal} {arXiv preprint arXiv:2510.06313}\ } (\bibinfo {year} {2025})}\BibitemShut {NoStop}%
\bibitem [{\citenamefont {Thaller}(2013)}]{thaller2013dirac}%
  \BibitemOpen
  \bibfield  {author} {\bibinfo {author} {\bibfnamefont {B.}~\bibnamefont {Thaller}},\ }\href@noop {} {\emph {\bibinfo {title} {The Dirac Equation}}}\ (\bibinfo  {publisher} {Springer Science \& Business Media},\ \bibinfo {year} {2013})\BibitemShut {NoStop}%
\bibitem [{\citenamefont {Nieves}\ and\ \citenamefont {Pal}(2004)}]{Nieves2004}%
  \BibitemOpen
  \bibfield  {author} {\bibinfo {author} {\bibfnamefont {J.~F.}\ \bibnamefont {Nieves}}\ and\ \bibinfo {author} {\bibfnamefont {P.~B.}\ \bibnamefont {Pal}},\ }\href {https://doi.org/10.1119/1.1757445} {\bibfield  {journal} {\bibinfo  {journal} {American Journal of Physics}\ }\textbf {\bibinfo {volume} {72}},\ \bibinfo {pages} {1100} (\bibinfo {year} {2004})}\BibitemShut {NoStop}%
\bibitem [{\citenamefont {Sigrist}\ and\ \citenamefont {Ueda}(1991)}]{Sigrist1991}%
  \BibitemOpen
  \bibfield  {author} {\bibinfo {author} {\bibfnamefont {M.}~\bibnamefont {Sigrist}}\ and\ \bibinfo {author} {\bibfnamefont {K.}~\bibnamefont {Ueda}},\ }\href {\doibase 10.1103/RevModPhys.63.239} {\bibfield  {journal} {\bibinfo  {journal} {Rev. Mod. Phys.}\ }\textbf {\bibinfo {volume} {63}},\ \bibinfo {pages} {239} (\bibinfo {year} {1991})}\BibitemShut {NoStop}%
\bibitem [{\citenamefont {Abanov}\ \emph {et~al.}(2001{\natexlab{a}})\citenamefont {Abanov}, \citenamefont {Chubukov},\ and\ \citenamefont {Finkel'stein}}]{abanov2001}%
  \BibitemOpen
  \bibfield  {author} {\bibinfo {author} {\bibfnamefont {A.}~\bibnamefont {Abanov}}, \bibinfo {author} {\bibfnamefont {A.~V.}\ \bibnamefont {Chubukov}}, \ and\ \bibinfo {author} {\bibfnamefont {A.~M.}\ \bibnamefont {Finkel'stein}},\ }\href {\doibase 10.1209/epl/i2001-00266-0} {\bibfield  {journal} {\bibinfo  {journal} {Europhysics Letters}\ }\textbf {\bibinfo {volume} {54}},\ \bibinfo {pages} {488} (\bibinfo {year} {2001}{\natexlab{a}})}\BibitemShut {NoStop}%
\bibitem [{\citenamefont {Abanov}\ \emph {et~al.}(2001{\natexlab{b}})\citenamefont {Abanov}, \citenamefont {Chubukov},\ and\ \citenamefont {Schmalian}}]{abanov2001b}%
  \BibitemOpen
  \bibfield  {author} {\bibinfo {author} {\bibfnamefont {A.}~\bibnamefont {Abanov}}, \bibinfo {author} {\bibfnamefont {A.~V.}\ \bibnamefont {Chubukov}}, \ and\ \bibinfo {author} {\bibfnamefont {J.}~\bibnamefont {Schmalian}},\ }\href {\doibase 10.1209/epl/i2001-00425-9} {\bibfield  {journal} {\bibinfo  {journal} {Europhysics Letters}\ }\textbf {\bibinfo {volume} {55}},\ \bibinfo {pages} {369} (\bibinfo {year} {2001}{\natexlab{b}})}\BibitemShut {NoStop}%
\bibitem [{\citenamefont {Chubukov}\ and\ \citenamefont {Schmalian}(2005)}]{ChubukovJS2005}%
  \BibitemOpen
  \bibfield  {author} {\bibinfo {author} {\bibfnamefont {A.~V.}\ \bibnamefont {Chubukov}}\ and\ \bibinfo {author} {\bibfnamefont {J.}~\bibnamefont {Schmalian}},\ }\href {\doibase 10.1103/PhysRevB.72.174520} {\bibfield  {journal} {\bibinfo  {journal} {Phys. Rev. B}\ }\textbf {\bibinfo {volume} {72}},\ \bibinfo {pages} {174520} (\bibinfo {year} {2005})}\BibitemShut {NoStop}%
\bibitem [{\citenamefont {Abanov}\ and\ \citenamefont {Chubukov}(2020)}]{abanov2020-I}%
  \BibitemOpen
  \bibfield  {author} {\bibinfo {author} {\bibfnamefont {A.}~\bibnamefont {Abanov}}\ and\ \bibinfo {author} {\bibfnamefont {A.~V.}\ \bibnamefont {Chubukov}},\ }\href {\doibase 10.1103/PhysRevB.102.024524} {\bibfield  {journal} {\bibinfo  {journal} {Phys. Rev. B}\ }\textbf {\bibinfo {volume} {102}},\ \bibinfo {pages} {024524} (\bibinfo {year} {2020})}\BibitemShut {NoStop}%
\bibitem [{\citenamefont {Wu}\ \emph {et~al.}(2020{\natexlab{a}})\citenamefont {Wu}, \citenamefont {Abanov}, \citenamefont {Wang},\ and\ \citenamefont {Chubukov}}]{abanov2020-II}%
  \BibitemOpen
  \bibfield  {author} {\bibinfo {author} {\bibfnamefont {Y.-M.}\ \bibnamefont {Wu}}, \bibinfo {author} {\bibfnamefont {A.}~\bibnamefont {Abanov}}, \bibinfo {author} {\bibfnamefont {Y.}~\bibnamefont {Wang}}, \ and\ \bibinfo {author} {\bibfnamefont {A.~V.}\ \bibnamefont {Chubukov}},\ }\href {\doibase 10.1103/PhysRevB.102.024525} {\bibfield  {journal} {\bibinfo  {journal} {Phys. Rev. B}\ }\textbf {\bibinfo {volume} {102}},\ \bibinfo {pages} {024525} (\bibinfo {year} {2020}{\natexlab{a}})}\BibitemShut {NoStop}%
\bibitem [{\citenamefont {Wu}\ \emph {et~al.}(2020{\natexlab{b}})\citenamefont {Wu}, \citenamefont {Abanov},\ and\ \citenamefont {Chubukov}}]{wu2020-III}%
  \BibitemOpen
  \bibfield  {author} {\bibinfo {author} {\bibfnamefont {Y.-M.}\ \bibnamefont {Wu}}, \bibinfo {author} {\bibfnamefont {A.}~\bibnamefont {Abanov}}, \ and\ \bibinfo {author} {\bibfnamefont {A.~V.}\ \bibnamefont {Chubukov}},\ }\href {\doibase 10.1103/PhysRevB.102.094516} {\bibfield  {journal} {\bibinfo  {journal} {Phys. Rev. B}\ }\textbf {\bibinfo {volume} {102}},\ \bibinfo {pages} {094516} (\bibinfo {year} {2020}{\natexlab{b}})}\BibitemShut {NoStop}%
\bibitem [{\citenamefont {Wu}\ \emph {et~al.}(2021{\natexlab{a}})\citenamefont {Wu}, \citenamefont {Zhang}, \citenamefont {Abanov},\ and\ \citenamefont {Chubukov}}]{wu2021-IV}%
  \BibitemOpen
  \bibfield  {author} {\bibinfo {author} {\bibfnamefont {Y.-M.}\ \bibnamefont {Wu}}, \bibinfo {author} {\bibfnamefont {S.-S.}\ \bibnamefont {Zhang}}, \bibinfo {author} {\bibfnamefont {A.}~\bibnamefont {Abanov}}, \ and\ \bibinfo {author} {\bibfnamefont {A.~V.}\ \bibnamefont {Chubukov}},\ }\href {\doibase 10.1103/PhysRevB.103.024522} {\bibfield  {journal} {\bibinfo  {journal} {Phys. Rev. B}\ }\textbf {\bibinfo {volume} {103}},\ \bibinfo {pages} {024522} (\bibinfo {year} {2021}{\natexlab{a}})}\BibitemShut {NoStop}%
\bibitem [{\citenamefont {Wu}\ \emph {et~al.}(2021{\natexlab{b}})\citenamefont {Wu}, \citenamefont {Zhang}, \citenamefont {Abanov},\ and\ \citenamefont {Chubukov}}]{wu2021-V}%
  \BibitemOpen
  \bibfield  {author} {\bibinfo {author} {\bibfnamefont {Y.-M.}\ \bibnamefont {Wu}}, \bibinfo {author} {\bibfnamefont {S.-S.}\ \bibnamefont {Zhang}}, \bibinfo {author} {\bibfnamefont {A.}~\bibnamefont {Abanov}}, \ and\ \bibinfo {author} {\bibfnamefont {A.~V.}\ \bibnamefont {Chubukov}},\ }\href {\doibase 10.1103/PhysRevB.103.184508} {\bibfield  {journal} {\bibinfo  {journal} {Phys. Rev. B}\ }\textbf {\bibinfo {volume} {103}},\ \bibinfo {pages} {184508} (\bibinfo {year} {2021}{\natexlab{b}})}\BibitemShut {NoStop}%
\bibitem [{\citenamefont {Ojaj{\"a}rvi}\ \emph {et~al.}(2024)\citenamefont {Ojaj{\"a}rvi}, \citenamefont {Chubukov}, \citenamefont {Lee}, \citenamefont {Garst},\ and\ \citenamefont {Schmalian}}]{Ojajarvi2024}%
  \BibitemOpen
  \bibfield  {author} {\bibinfo {author} {\bibfnamefont {R.}~\bibnamefont {Ojaj{\"a}rvi}}, \bibinfo {author} {\bibfnamefont {A.~V.}\ \bibnamefont {Chubukov}}, \bibinfo {author} {\bibfnamefont {Y.-C.}\ \bibnamefont {Lee}}, \bibinfo {author} {\bibfnamefont {M.}~\bibnamefont {Garst}}, \ and\ \bibinfo {author} {\bibfnamefont {J.}~\bibnamefont {Schmalian}},\ }\href {https://doi.org/10.1038/s41535-024-00717-4} {\bibfield  {journal} {\bibinfo  {journal} {npj Quantum Materials}\ }\textbf {\bibinfo {volume} {9}},\ \bibinfo {pages} {105} (\bibinfo {year} {2024})}\BibitemShut {NoStop}%
\bibitem [{\citenamefont {Wang}\ and\ \citenamefont {Chubukov}(2025)}]{Wang_Chub_2025}%
  \BibitemOpen
  \bibfield  {author} {\bibinfo {author} {\bibfnamefont {Y.}~\bibnamefont {Wang}}\ and\ \bibinfo {author} {\bibfnamefont {A.~V.}\ \bibnamefont {Chubukov}},\ }\href {\doibase 10.1103/ckyl-flxm} {\bibfield  {journal} {\bibinfo  {journal} {Phys. Rev. B}\ }\textbf {\bibinfo {volume} {111}},\ \bibinfo {pages} {214514} (\bibinfo {year} {2025})}\BibitemShut {NoStop}%
\bibitem [{\citenamefont {Kaplan}\ \emph {et~al.}(2009)\citenamefont {Kaplan}, \citenamefont {Lee}, \citenamefont {Son},\ and\ \citenamefont {Stephanov}}]{Kaplan2009}%
  \BibitemOpen
  \bibfield  {author} {\bibinfo {author} {\bibfnamefont {D.~B.}\ \bibnamefont {Kaplan}}, \bibinfo {author} {\bibfnamefont {J.-W.}\ \bibnamefont {Lee}}, \bibinfo {author} {\bibfnamefont {D.~T.}\ \bibnamefont {Son}}, \ and\ \bibinfo {author} {\bibfnamefont {M.~A.}\ \bibnamefont {Stephanov}},\ }\href {\doibase 10.1103/PhysRevD.80.125005} {\bibfield  {journal} {\bibinfo  {journal} {Phys. Rev. D}\ }\textbf {\bibinfo {volume} {80}},\ \bibinfo {pages} {125005} (\bibinfo {year} {2009})}\BibitemShut {NoStop}%
\bibitem [{\citenamefont {Christos}\ \emph {et~al.}(2023)\citenamefont {Christos}, \citenamefont {Sachdev},\ and\ \citenamefont {Scheurer}}]{Christos2023nodal}%
  \BibitemOpen
  \bibfield  {author} {\bibinfo {author} {\bibfnamefont {M.}~\bibnamefont {Christos}}, \bibinfo {author} {\bibfnamefont {S.}~\bibnamefont {Sachdev}}, \ and\ \bibinfo {author} {\bibfnamefont {M.~S.}\ \bibnamefont {Scheurer}},\ }\href {https://doi.org/10.1038/s41467-023-42471-4} {\bibfield  {journal} {\bibinfo  {journal} {Nature Communications}\ }\textbf {\bibinfo {volume} {14}},\ \bibinfo {pages} {7134} (\bibinfo {year} {2023})}\BibitemShut {NoStop}%
\bibitem [{\citenamefont {Berezinskii}(1971)}]{Berezinskii1971}%
  \BibitemOpen
  \bibfield  {author} {\bibinfo {author} {\bibfnamefont {V.~L.}\ \bibnamefont {Berezinskii}},\ }\href@noop {} {\bibfield  {journal} {\bibinfo  {journal} {Sov. Phys. JETP}\ }\textbf {\bibinfo {volume} {32}},\ \bibinfo {pages} {493} (\bibinfo {year} {1971})}\BibitemShut {NoStop}%
\bibitem [{\citenamefont {Kosterlitz}\ and\ \citenamefont {Thouless}(1973)}]{Kosterlitz1973}%
  \BibitemOpen
  \bibfield  {author} {\bibinfo {author} {\bibfnamefont {J.~M.}\ \bibnamefont {Kosterlitz}}\ and\ \bibinfo {author} {\bibfnamefont {D.~J.}\ \bibnamefont {Thouless}},\ }\href {https://dx.doi.org/10.1088/0022-3719/6/7/010} {\bibfield  {journal} {\bibinfo  {journal} {J. Phys. C: Solid State Phys. 6 1181}\ }\textbf {\bibinfo {volume} {6}},\ \bibinfo {pages} {1181} (\bibinfo {year} {1973})}\BibitemShut {NoStop}%
\bibitem [{\citenamefont {Halperin}\ and\ \citenamefont {Nelson}(1979)}]{Halperin1979}%
  \BibitemOpen
  \bibfield  {author} {\bibinfo {author} {\bibfnamefont {B.}~\bibnamefont {Halperin}}\ and\ \bibinfo {author} {\bibfnamefont {D.~R.}\ \bibnamefont {Nelson}},\ }\href {https://doi.org/10.1007/BF00116988} {\bibfield  {journal} {\bibinfo  {journal} {Journal of Low Temperature Physics}\ }\textbf {\bibinfo {volume} {36}},\ \bibinfo {pages} {599} (\bibinfo {year} {1979})}\BibitemShut {NoStop}%
\bibitem [{\citenamefont {Raines}\ \emph {et~al.}(2024)\citenamefont {Raines}, \citenamefont {Zhang},\ and\ \citenamefont {Chubukov}}]{Raines2024}%
  \BibitemOpen
  \bibfield  {author} {\bibinfo {author} {\bibfnamefont {Z.~M.}\ \bibnamefont {Raines}}, \bibinfo {author} {\bibfnamefont {S.-S.}\ \bibnamefont {Zhang}}, \ and\ \bibinfo {author} {\bibfnamefont {A.~V.}\ \bibnamefont {Chubukov}},\ }\href {\doibase 10.1103/PhysRevB.109.144505} {\bibfield  {journal} {\bibinfo  {journal} {Phys. Rev. B}\ }\textbf {\bibinfo {volume} {109}},\ \bibinfo {pages} {144505} (\bibinfo {year} {2024})}\BibitemShut {NoStop}%
\bibitem [{\citenamefont {Ferrell}\ and\ \citenamefont {Glover}(1958)}]{Ferrell1958}%
  \BibitemOpen
  \bibfield  {author} {\bibinfo {author} {\bibfnamefont {R.~A.}\ \bibnamefont {Ferrell}}\ and\ \bibinfo {author} {\bibfnamefont {R.~E.}\ \bibnamefont {Glover}},\ }\href {\doibase 10.1103/PhysRev.109.1398} {\bibfield  {journal} {\bibinfo  {journal} {Phys. Rev.}\ }\textbf {\bibinfo {volume} {109}},\ \bibinfo {pages} {1398} (\bibinfo {year} {1958})}\BibitemShut {NoStop}%
\bibitem [{\citenamefont {Tinkham}\ and\ \citenamefont {Ferrell}(1959)}]{Tinkham1959}%
  \BibitemOpen
  \bibfield  {author} {\bibinfo {author} {\bibfnamefont {M.}~\bibnamefont {Tinkham}}\ and\ \bibinfo {author} {\bibfnamefont {R.~A.}\ \bibnamefont {Ferrell}},\ }\href {\doibase 10.1103/PhysRevLett.2.331} {\bibfield  {journal} {\bibinfo  {journal} {Phys. Rev. Lett.}\ }\textbf {\bibinfo {volume} {2}},\ \bibinfo {pages} {331} (\bibinfo {year} {1959})}\BibitemShut {NoStop}%
\bibitem [{\citenamefont {Sabio}\ \emph {et~al.}(2008)\citenamefont {Sabio}, \citenamefont {Nilsson},\ and\ \citenamefont {Castro~Neto}}]{Sabio2008}%
  \BibitemOpen
  \bibfield  {author} {\bibinfo {author} {\bibfnamefont {J.}~\bibnamefont {Sabio}}, \bibinfo {author} {\bibfnamefont {J.}~\bibnamefont {Nilsson}}, \ and\ \bibinfo {author} {\bibfnamefont {A.~H.}\ \bibnamefont {Castro~Neto}},\ }\href {\doibase 10.1103/PhysRevB.78.075410} {\bibfield  {journal} {\bibinfo  {journal} {Phys. Rev. B}\ }\textbf {\bibinfo {volume} {78}},\ \bibinfo {pages} {075410} (\bibinfo {year} {2008})}\BibitemShut {NoStop}%
\bibitem [{\citenamefont {Fritz}\ \emph {et~al.}(2008)\citenamefont {Fritz}, \citenamefont {Schmalian}, \citenamefont {M\"uller},\ and\ \citenamefont {Sachdev}}]{Fritz2008}%
  \BibitemOpen
  \bibfield  {author} {\bibinfo {author} {\bibfnamefont {L.}~\bibnamefont {Fritz}}, \bibinfo {author} {\bibfnamefont {J.}~\bibnamefont {Schmalian}}, \bibinfo {author} {\bibfnamefont {M.}~\bibnamefont {M\"uller}}, \ and\ \bibinfo {author} {\bibfnamefont {S.}~\bibnamefont {Sachdev}},\ }\href {\doibase 10.1103/PhysRevB.78.085416} {\bibfield  {journal} {\bibinfo  {journal} {Phys. Rev. B}\ }\textbf {\bibinfo {volume} {78}},\ \bibinfo {pages} {085416} (\bibinfo {year} {2008})}\BibitemShut {NoStop}%
\bibitem [{\citenamefont {Inkof}\ \emph {et~al.}(2020)\citenamefont {Inkof}, \citenamefont {K{\"u}ppers}, \citenamefont {Link}, \citenamefont {Gout{\'e}raux},\ and\ \citenamefont {Schmalian}}]{Inkof2020}%
  \BibitemOpen
  \bibfield  {author} {\bibinfo {author} {\bibfnamefont {G.~A.}\ \bibnamefont {Inkof}}, \bibinfo {author} {\bibfnamefont {J.}~\bibnamefont {K{\"u}ppers}}, \bibinfo {author} {\bibfnamefont {J.~M.}\ \bibnamefont {Link}}, \bibinfo {author} {\bibfnamefont {B.}~\bibnamefont {Gout{\'e}raux}}, \ and\ \bibinfo {author} {\bibfnamefont {J.}~\bibnamefont {Schmalian}},\ }\href {https://doi.org/10.1007/JHEP11(2020)088} {\bibfield  {journal} {\bibinfo  {journal} {Journal of High Energy Physics}\ }\textbf {\bibinfo {volume} {88}},\ \bibinfo {pages} {1} (\bibinfo {year} {2020})}\BibitemShut {NoStop}%
\bibitem [{\citenamefont {Liu}\ \emph {et~al.}(2010)\citenamefont {Liu}, \citenamefont {Qi}, \citenamefont {Zhang}, \citenamefont {Dai}, \citenamefont {Fang},\ and\ \citenamefont {Zhang}}]{Liu2010}%
  \BibitemOpen
  \bibfield  {author} {\bibinfo {author} {\bibfnamefont {C.-X.}\ \bibnamefont {Liu}}, \bibinfo {author} {\bibfnamefont {X.-L.}\ \bibnamefont {Qi}}, \bibinfo {author} {\bibfnamefont {H.}~\bibnamefont {Zhang}}, \bibinfo {author} {\bibfnamefont {X.}~\bibnamefont {Dai}}, \bibinfo {author} {\bibfnamefont {Z.}~\bibnamefont {Fang}}, \ and\ \bibinfo {author} {\bibfnamefont {S.-C.}\ \bibnamefont {Zhang}},\ }\href {\doibase 10.1103/PhysRevB.82.045122} {\bibfield  {journal} {\bibinfo  {journal} {Phys. Rev. B}\ }\textbf {\bibinfo {volume} {82}},\ \bibinfo {pages} {045122} (\bibinfo {year} {2010})}\BibitemShut {NoStop}%
\bibitem [{\citenamefont {Palle}\ and\ \citenamefont {Schmalian}(2024)}]{Palle2024b}%
  \BibitemOpen
  \bibfield  {author} {\bibinfo {author} {\bibfnamefont {G.}~\bibnamefont {Palle}}\ and\ \bibinfo {author} {\bibfnamefont {J.}~\bibnamefont {Schmalian}},\ }\href {\doibase 10.1103/PhysRevB.110.104516} {\bibfield  {journal} {\bibinfo  {journal} {Phys. Rev. B}\ }\textbf {\bibinfo {volume} {110}},\ \bibinfo {pages} {104516} (\bibinfo {year} {2024})}\BibitemShut {NoStop}%
\bibitem [{\citenamefont {Shi}\ \emph {et~al.}(2023)\citenamefont {Shi}, \citenamefont {Else}, \citenamefont {Goldman},\ and\ \citenamefont {Senthil}}]{Shi2023loop}%
  \BibitemOpen
  \bibfield  {author} {\bibinfo {author} {\bibfnamefont {Z.~D.}\ \bibnamefont {Shi}}, \bibinfo {author} {\bibfnamefont {D.~V.}\ \bibnamefont {Else}}, \bibinfo {author} {\bibfnamefont {H.}~\bibnamefont {Goldman}}, \ and\ \bibinfo {author} {\bibfnamefont {T.}~\bibnamefont {Senthil}},\ }\href {\doibase 10.21468/SciPostPhys.14.5.113} {\bibfield  {journal} {\bibinfo  {journal} {SciPost Phys.}\ }\textbf {\bibinfo {volume} {14}},\ \bibinfo {pages} {113} (\bibinfo {year} {2023})}\BibitemShut {NoStop}%
\bibitem [{\citenamefont {Palle}\ \emph {et~al.}(2024)\citenamefont {Palle}, \citenamefont {Ojaj{\"a}rvi}, \citenamefont {Fernandes},\ and\ \citenamefont {Schmalian}}]{Palle2024}%
  \BibitemOpen
  \bibfield  {author} {\bibinfo {author} {\bibfnamefont {G.}~\bibnamefont {Palle}}, \bibinfo {author} {\bibfnamefont {R.}~\bibnamefont {Ojaj{\"a}rvi}}, \bibinfo {author} {\bibfnamefont {R.~M.}\ \bibnamefont {Fernandes}}, \ and\ \bibinfo {author} {\bibfnamefont {J.}~\bibnamefont {Schmalian}},\ }\href {https://www.science.org/doi/full/10.1126/sciadv.adn3662} {\bibfield  {journal} {\bibinfo  {journal} {Science Advances}\ }\textbf {\bibinfo {volume} {10}},\ \bibinfo {pages} {eadn3662} (\bibinfo {year} {2024})}\BibitemShut {NoStop}%
\bibitem [{\citenamefont {Schultz}\ \emph {et~al.}(2025)\citenamefont {Schultz}, \citenamefont {Palle}, \citenamefont {Kim}, \citenamefont {Fernandes},\ and\ \citenamefont {Schmalian}}]{Schultz2025}%
  \BibitemOpen
  \bibfield  {author} {\bibinfo {author} {\bibfnamefont {D.~J.}\ \bibnamefont {Schultz}}, \bibinfo {author} {\bibfnamefont {G.}~\bibnamefont {Palle}}, \bibinfo {author} {\bibfnamefont {Y.~B.}\ \bibnamefont {Kim}}, \bibinfo {author} {\bibfnamefont {R.~M.}\ \bibnamefont {Fernandes}}, \ and\ \bibinfo {author} {\bibfnamefont {J.}~\bibnamefont {Schmalian}},\ }\href {https://arxiv.org/abs/2507.16892} {\bibfield  {journal} {\bibinfo  {journal} {arXiv preprint arXiv:2507.16892}\ } (\bibinfo {year} {2025})}\BibitemShut {NoStop}%
\end{thebibliography}%

\end{document}